\documentclass[twocolumn,pra,aps,preprintnumbers,showpacs,superscriptaddress]{revtex4-1}

\usepackage{latexsym}
\usepackage{amssymb}
\usepackage{amsmath}
\usepackage{amsthm}
\usepackage{amsfonts}
\usepackage{ifthen}
\usepackage{revsymb}
\usepackage{yfonts}
\usepackage{graphicx}
\usepackage{algpseudocode}
\usepackage{bbm}
\usepackage{multirow}
\usepackage{gensymb}

\usepackage[colorlinks = true]{hyperref}

\usepackage{xcolor}
\definecolor{darkred}  {rgb}{0.5,0,0}
\definecolor{darkblue} {rgb}{0,0,0.5}
\definecolor{darkgreen}{rgb}{0,0.5,0}

\hypersetup{
  urlcolor   = blue,         
  linkcolor  = darkblue,     
  citecolor  = darkgreen,    
  filecolor  = darkred       
}

\newcommand{\calA}{{\cal A }}

\newcommand{\calC}{{\cal C }}

\newcommand{\calH}{{\cal H }}
\newcommand{\calF}{{\cal F }}

\newcommand{\calE}{{\cal E }}

\newcommand{\calZ}{{\cal Z }}
\newcommand{\calX}{{\cal X }}
\newcommand{\calY}{{\cal Y }}
\newcommand{\calO}{{\cal O }}
\newcommand{\calP}{{\cal P }}
\newcommand{\calQ}{{\cal Q }}
\newcommand{\calU}{{\cal U }}

\newcommand{\calS}{{\cal S }}

\newcommand{\be}{\begin{equation}}
\newcommand{\ee}{\end{equation}}
\newcommand{\bq}{\begin{eqnarray}}
\newcommand{\eq}{\end{eqnarray}}
\newcommand{\bea}{\begin{eqnarray}}
\newcommand{\eea}{\end{eqnarray}}
\newcommand{\ba}{\begin{align}}
\newcommand{\ea}{\end{align}}

\newcommand{\bC}{\mathbbm{C}}
\newcommand{\bZ}{\mathbbm{Z}}

\newcommand{\cO}{\mathcal O}

\newcommand{\cP}{\mathcal P}

\newcommand{\Sp}{\,\,\,\,\,\,}
\newcommand{\no}{\nonumber\\}

\definecolor{mygray}{gray}{0.6}
\newcommand{\mgr}[1]{{\color{mygray} #1}}

\begin{document}

\title{Tapering off qubits to simulate fermionic Hamiltonians}

\author{Sergey \surname{Bravyi}}
\affiliation{IBM T.J.  Watson  Research  Center,  Yorktown  Heights,  NY 10598,  USA}
\author{Jay  M. \surname{Gambetta}}
\affiliation{IBM T.J.  Watson  Research  Center,  Yorktown  Heights,  NY 10598,  USA}
\author{Antonio \surname{Mezzacapo}}
\affiliation{IBM T.J.  Watson  Research  Center,  Yorktown  Heights,  NY 10598,  USA}
\author{Kristan \surname{Temme}}
\affiliation{IBM T.J.  Watson  Research  Center,  Yorktown  Heights,  NY 10598,  USA}

\date{\today}

\begin{abstract}
We discuss encodings of fermionic many-body systems by qubits in the presence of symmetries.
Such encodings eliminate redundant degrees of freedom in a way that preserves a simple structure
of the system Hamiltonian enabling  quantum simulations with fewer qubits. 
First we consider $U(1)$ symmetry describing   the particle number conservation. 
Using a previously known encoding based on the  first quantization method
a system of $M$ fermi  modes with  $N$ particles
can be simulated on a quantum computer with $Q=N\log_2{(M)}$ qubits.
We propose a new version of this encoding tailored to  
variational quantum algorithms. Also we show how to improve  sparsity of the simulator Hamiltonian
using  orthogonal arrays.
Next we consider encodings based on the second quantization method.
It is shown that encodings with a given filling fraction  $\nu=N/M$ and a qubit-per-mode ratio $\eta=Q/M<1$
can be constructed from  efficiently decodable classical LDPC codes with the 
relative distance $2\nu$ and the
encoding rate $1-\eta$. 
A   family of codes based on high-girth bipartite graphs
is discussed.  Graph-based encodings  eliminate roughly $M/N$ qubits.
Finally we consider $\mathbb{Z}_2$ symmetries,
and show how to eliminate qubits using previously known encodings,
illustrating the technique for simple molecular-type Hamiltonians. 
\end{abstract}
\maketitle

\section{Introduction}
Quantum information processing holds the promise of solving some of the problems 
that are  deemed too  challenging for conventional computers.
One important problem in this category is the simulation 
of strongly interacting fermionic systems, in the context of quantum chemistry or material science.
A natural application for a quantum computer would be preparing 
low-energy states  and estimating the ground energy of a  fermionic Hamiltonian.
Several methods have been  proposed in the literature to accomplish this task,
for example, preparation of a good trial state followed by the 
quantum phase estimation~\cite{AbramsLloyd1,aspuru2005simulated} or
state preparation by the adiabatic evolution~\cite{farhi2000quantum,babbush2014adiabatic}.
These methods however require a universal quantum computer
capable of implementing very long circuits, exceeding
the state-of-the-art demonstrations by many orders of magnitude~\cite{wecker2015solving}.
Alternative methods that could be more viable in the near future
are  variational quantum algorithms~\cite{peruzzo2014variational,wecker2015progress,mcclean2016theory,Li2016efficient}.
Such algorithms minimize the energy  of a target fermionic Hamiltonian
over a class of trial states that can be 
prepared on a given quantum hardware by varying control parameters. 

Since the basic units of a quantum computer are qubits rather than fermi modes,
any simulation method relies on a certain encoding of fermionic degrees of freedom
by qubits~\cite{Wigner1928,BK2002,verstraete2005mapping,seeley2012bravyi,Tranter15,Moll2015,whitfield2016local}.
A choice of a good encoding is important as it may affect  both
the number of qubits and the running time of a simulation algorithm.
Here we propose  encodings tailored to variational quantum
algorithms and fermionic Hamiltonians that possess symmetries such as
the particle number conservation.  The presence of symmetries
allows one to restrict the simulation to an eigenspace of the symmetry operator
whereby reducing the number of qubits that encode a fermionic system.
In several specific examples, such as the hydrogen molecule and
the Fermi-Hubbdard model  it has been observed that some qubits can
indeed  be removed from the simulation without loss of information~\cite{o2015scalable,Moll2015}.
Importantly, the removal of qubits in these examples preserves a simple structure 
of the encoded Hamiltonian  enabling efficient simulations with fewer qubits. 
This motivates the question of how to generalize these methods and how to eliminate redundant qubits
in a  computationally efficient manner for larger systems.

To address these questions   let us first give a more precise notion of a
simulation.  We shall describe  a fermionic system to be simulated  by
a target Hamiltonian $H_{tgt}$ composed of one- and two-body operators such as those describing hopping, chemical potential,
and two-particle interactions, 
\be
\label{Htgt}
H_{tgt} =\sum_{\alpha, \beta=1}^M t_{\alpha \beta} \, a^\dag_{\alpha} a_{\beta} +  \sum_{\alpha, \beta, \gamma, \delta = 1}^M u_{\alpha \beta \gamma \delta}\, a^\dag_{\alpha} a^\dag_{\beta} a_{\gamma} a_{\delta}.
\ee 
Here $M$ is the number of fermi modes,
 $a^\dag_\alpha$  and $a_\alpha$ are  creation and annihilation operators for a mode $\alpha$,
and $t_{\alpha \beta},u_{\alpha \beta \gamma \delta}$ are complex coefficients
such that $t_{\beta\alpha}=t_{\alpha\beta}^*$ and $u_{\alpha \beta \gamma \delta}=u_{\delta\gamma\beta\alpha}^*$.
Leaving aside superconductivity and relativistic effects, all natural fermionic Hamiltonians
have the above form.
Since each term in $H_{tgt}$ has equal number of creation and annihilation operators,
$H_{tgt}$ commutes with the particle number operator 
$\hat{N}\equiv \sum_{\alpha=1}^M a_\alpha^\dag a_\alpha$.
We shall assume  that the system contains a fixed number of particles $N$.
For example, $N$ could be the number of valence electrons in a molecule.
Define a  target  Hilbert space $\calH_{tgt}$
as the $N$-particle subspace 
 spanned by all states $|\phi\rangle$ of $M$ fermi modes satisfying 
$\hat{N}|\phi\rangle=N|\phi\rangle$. Without loss of generality, $N\le M/2$
(otherwise consider holes instead of particles). 
Our goal is to estimate the minimum energy of $H_{tgt}$ 
restricted to the $N$-particle subspace $\calH_{tgt}$.
Below we shall often identify $H_{tgt}$ and the restriction
of $H_{tgt}$ onto the subspace $\calH_{tgt}$.
We shall choose the energy
scale such that all coefficients in Eq.~(\ref{Htgt}) have magnitude at most one.

Let us now formally define an encoding of fermionic degrees of freedom by qubits. 
Following Ref.~\cite{TIM2014}, 
we shall describe such encoding  as an isometry
$\calE \, : \, \calH_{tgt}\to \calH_{sim}$, where 
$\calH_{sim}=(\bC^2)^{\otimes Q}$ is the Hilbert space of $Q$ qubits.
A state of the target system $|\phi\rangle \in \calH_{tgt}$ is identified
with a state $\calE|\phi\rangle$ of the simulator system.
Encoded states $\calE|\phi\rangle$ span a {\em codespace}
$\mathrm{Im}(\calE)\equiv \calE\cdot \calH_{tgt}$.

A Hamiltonian $H_{sim}$ describing a system of $Q$ qubits  is called a {\em simulator}
of $H_{tgt}$ if it satisfies two conditions. First, we require that
\begin{equation}
\label{simulation}
H_{sim} \calE = \calE H_{tgt}.
\end{equation}
In words, $H_{sim}$  must preserve the codespace  
and the restriction of $H_{sim}$ onto the codespace must be unitarily  equivalent to $H_{tgt}$.  
The action of $H_{sim}$ on the orthogonal complement of the codespace may be
arbitrary. Secondly, we require that the codespace contains a ground state 
of $H_{sim}$. This guarantees that  $H_{tgt}$ and $H_{sim}$ have
the same ground energy and their ground states coincide modulo the  encoding $\calE$.

Our goal is to construct encodings that require $Q<M$ qubits
and, at the same time, all  target Hamiltonians Eq.~(\ref{Htgt}) have 
sufficiently simple simulators. We shall say that a simulator Hamiltonian 
is {\em $r$-sparse} if 
\begin{equation}
\label{Hsim}
H_{sim}=\sum_{i=1}^r D_i,
\end{equation}
where each operator $D_i$ is
diagonal in some tensor product basis of $Q$ qubits.
This basis may depend on $i$. 
We require that matrix elements of $D_i$ are efficiently computable.
A family of  encodings $\calE$ as above is called sparse  if
there exist small constants $c,d$ such that  
any target Hamiltonian Eq.~(\ref{Htgt}) has an $r$-sparse simulator Eq.~(\ref{Hsim}) with $r\le M^c$
and $\|D_i\|\le M^d$ for all $M$. 
Sparse encodings are well-suited for applications in variational quantum
algorithms~\cite{peruzzo2014variational,wecker2015progress,mcclean2016theory,Li2016efficient}.
Indeed,  a basic subroutine of variational algorithms 
is estimating energy $\langle \psi |H_{sim}|\psi\rangle$ of a given trial state
$\psi\in \calH_{sim}$ that can be prepared on the available quantum hardware. 
One can estimate the expectation value $e_i\equiv \langle \psi |D_i|\psi\rangle$ 
by preparing the trial state $\psi$, performing a local change of basis
in each qubit such that $D_i$ becomes diagonal
in the standard $|0\rangle$, $|1\rangle$ basis, and then measuring each qubit.
Performing a sequence of such measurements 
with a freshly prepared trial  state $\psi$ for each term $D_i$
gives an estimate of the energy $\langle \psi|H_{sim}|\psi\rangle=\sum_{i=1}^r e_i$.

\subsection*{Summary of results}

First  assume that  the number of particles is small such that
$N\log_2{(M)}<M$. We expect that this regime may be relevant for 
high-accuracy quantum chemistry calculations with large basis sets. 
We construct a sparse encoding with $Q=N\log_2{(M)}$ qubits such that 
any target Hamiltonian Eq.~(\ref{Htgt})  has a  simulator Eq.~(\ref{Hsim}) with sparsity 
\begin{equation}
\label{result1}
r\le 9 M^{3.17}.
\end{equation}
We also give a $poly(M)$ upper bound on the norm of the terms $D_i$
although we do not expect this bound to be tight.
This encoding mostly follows ideas of Refs.~\cite{Toloui2013,Babbush2015exponentially} and relies 
on the  first-quantization method. 
We extend the results of Refs.~\cite{Toloui2013,Babbush2015exponentially} in two respects.
First we show how to improve the sparsity of $H_{sim}$
using  orthogonal arrays~\cite{Hedayat2012orthogonal}.
Such arrays have been previously used
for quantum simulations and dynamical decoupling~\cite{Wocjan2002,Kern2005,Rotteler2006}
but their application in the context of variational quantum algorithms appears to be new. 
Secondly, we introduce a sparse Hamiltonian enforcing the anti-symmetric structure
of encoded states and compute the spectral gap of this Hamiltonian using
arguments based on the Schur duality~\cite{Goodman2000representations,bacon2005quantum,Bacon2006efficient}.
This allows us to bound the norm $\|D_i\|$.

Next consider encodings based on the second quantization method.
Let $\nu=N/M$ be the filling fraction and $\eta=Q/M$
be the desired qubit-per-mode ratio. 
We show  that sparse encodings with prescribed $\nu$, $\eta$ can be constructed from
classical error correcting codes with certain properties. Namely, we need 
binary linear codes that have a column-sparse parity check matrix,
relative distance $2\nu$,  and the encoding rate $1-\eta$.
It is known that the requisite  codes  can be constructed whenever 
\begin{equation}
\label{result1}
\nu<1/4 \quad  \mbox{and} \quad  \eta >h(2\nu),
\end{equation}
where $h(x)=-x\log_2{(x)} -(1-x)\log_2{(1-x)}$ is the binary Shannon entropy. 
For example, one can use  good LDPC codes~\cite{Gallager1962} whose parameters
approach the  Gilbert-Varshamov bound~\cite{Varshamov1957,MacWilliamsBook}.
Thus a constant fraction of qubits can be eliminated if the target system has a filling fraction 
$\nu<1/4$. The simulator Hamiltonian Eq.~(\ref{Hsim}) has sparsity $r$
proportional to the number
of non-zero coefficients $t_{\alpha\beta}$, $u_{\alpha\beta\gamma\delta}$ in 
the target Hamiltonian. Furthermore, $\|D_i\|\le 1$ for all $i$.

The bound Eq.~(\ref{result1}) is worse than what one could expect naively. Indeed, since 
$\calH_{tgt}$ has dimension  ${M\choose N}\approx 2^{M h(\nu)}$, 
the information-theoretic bound on the qubit-per-mode ratio 
is $\eta \ge h(\nu)$.  We leave as an open question whether sparse
encodings can achieve this bound.

The above  result  has one important caveat.
Namely,  we shall see that matrix elements of the simulator Hamiltonian 
can be computed efficiently only if the chosen code is efficiently decodable. In Appendix~\ref{app:decoding}
 we describe a brute force implementation
of the decoding algorithm that may be practical for small number of modes $M\le 50$.
Simulating larger systems may require LDPC codes that are both good and efficiently decodable.
Designing such codes is an active research
area, see Refs.~\cite{sipser1996expander,zemor2001expander,guruswami2005linear}.
 
 \begin{figure}[h]
\centerline{\includegraphics[height=3.75cm]{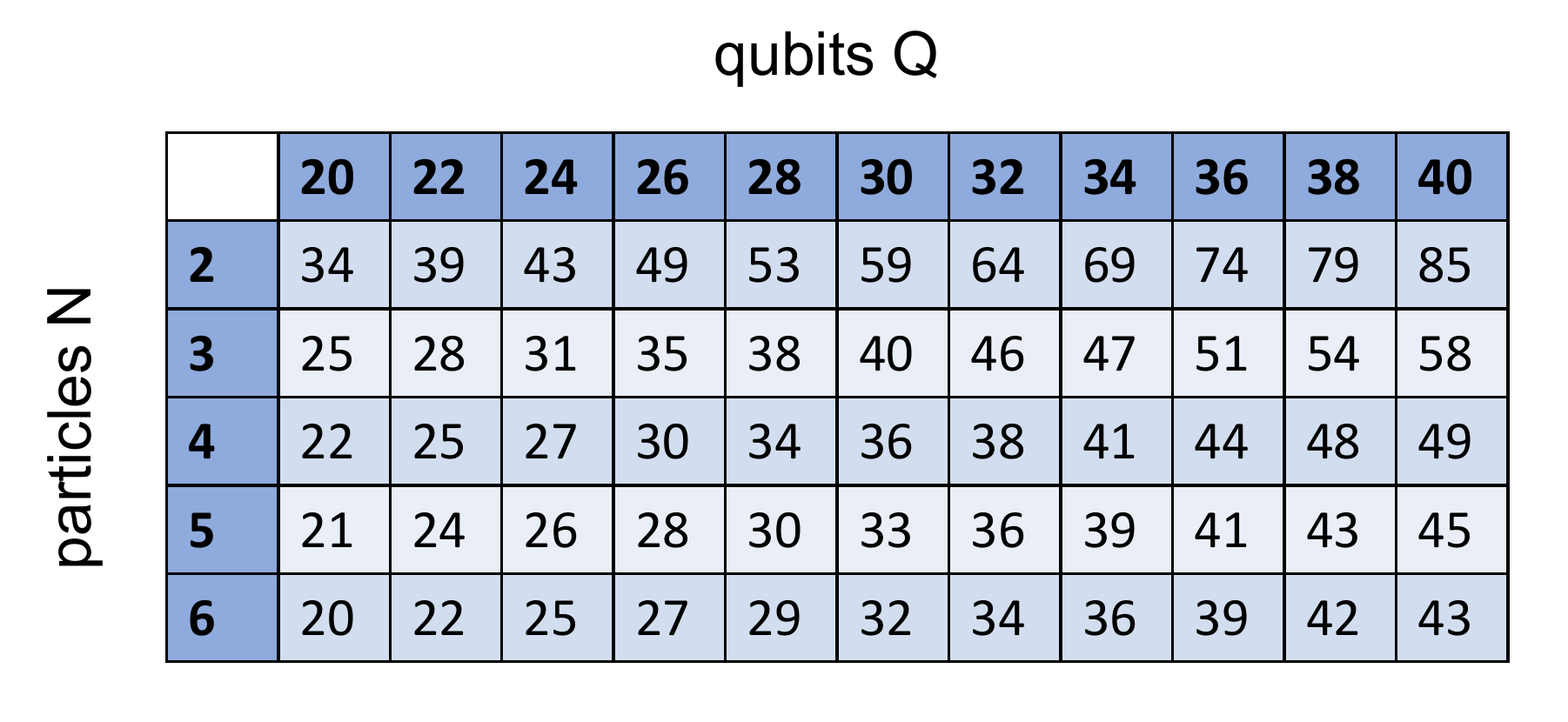}}
\caption{A lower bound on the number of fermionic modes $M$ that can be
simulated for fixed values of $N$ and $Q$ using the graph-based encodings. }
\label{figure:1}
\end{figure}

To  enable efficient decoding and improve sparseness of the simulator Hamiltonian
we consider  a  special class of LDPC codes associated with high-girth bipartite graphs.
For such encodings any two-body operator $a_\alpha^\dag a_\beta\pm a_\beta^\dag a_\alpha$
has a $2$-sparse simulator, while any four-body operator has $32$-sparse simulator.
Furthermore, matrix elements of the simulators can be computed in time $O(M^3)$. 
Graph-based encodings can eliminate $M/N$
qubits for $N\le M^{1/2}$. 
Figure~1 shows a numerically computed lower bound on the number
of modes $M$ that can be simulated for fixed values of $Q,N$
using the graph-based encodings.

Our third result concerns $\bZ_2$-symmetries such as the fermionic parity conservation. 
For concreteness, we consider a symmetry group $\bZ_2\times \bZ_2$ describing
parity conservation for electrons with a fixed spin orientation. Such symmetry is present, for example,
in the standard molecular electron Hamiltonians that neglect spin-spin and spin-orbit interactions.
The target Hilbert  space $\calH_{tgt}$ is chosen as a
subspace in which the number of electrons with a given spin
is fixed modulo two.
Using the second quantization method and both the parity and the binary tree encoding of Ref.~\cite{BK2002}
we construct a  simulator Hamiltonian Eq.~(\ref{Hsim}) that acts  on $Q=M-2$ qubits.
More generally, we give a systematic
method of detecting $\bZ_2$ symmetries in a given target Hamiltonian
and show how to construct sparse encodings that
eliminate one qubit for each $\bZ_2$ symmetry.
Our techniques are illustrated using  quantum chemistry Hamiltonians
describing simple molecules.

To summarize, we observed that the encoding  based on the first quantization method 
achieves the best performance in terms of the number of qubits that can be eliminated.
However, it is applicable only if the number of particles $N$ is relatively small.
Furthermore, the encoding does not take advantage of any structure present in 
the coefficients of $H_{tgt}$. 
For example,  $H_{sim}$ might have sparseness $9M^{3.17}$
even if $H_{tgt}$ has only $O(M)$ non-zero coefficients. In contrast,  encodings based on the second quantization method
eliminate fewer qubits but have broader applicability 
and  produce more sparse simulator
Hamiltonian such that the number of terms in $H_{tgt}$ and $H_{sim}$ are roughly the same
(up to a constant factor). 
 Which encoding should be preferred may depend on details
of the target system. 

We expect our results to find applications in the near-future experimental demonstrations of variational quantum algorithms.
In this context the number of qubits $Q$ is fixed  by the hardware constraints and may not be large enough to
simulate interesting molecules directly.  Combining the standard
variational algorithms~\cite{peruzzo2014variational,wecker2015progress,mcclean2016theory} with the encodings
described in this paper
may extend the range of molecules that can be simulated on a given  quantum hardware.
We leave as open questions whether sparse encoding can be constructed for
the filling fraction $\nu\ge 1/4$, what is the tradeoff between the 
parameters $M,N,Q$ and the encoding sparseness, 
 and how to generalize our techniques to other types of symmetries.

The rest of the paper is organized as follows. Section~\ref{sec:notations}
summarizes our notations.
Encodings based on the first quantization method are described
in Section~\ref{sec:small}. These encoding are applicable  if the number
of particles $N$ is sufficiently  small. 
Section~\ref{sec:LDPC}  shows how to construct sparse encodings
for a constant filling fraction $N/M$ using classical LDPC codes.
Encodings based on high-girth bipartite graphs
are described in Sections~\ref{sec:improved},\ref{sec:graph}.
Discrete  symmetries  and applications of our techniques
to small molecular-type Hamiltonians are discussed  in Sections~\ref{sec:parity},\ref{sec:multiple}.
Appendix~\ref{app:StandardMaps} summarizes the previously known encodings of fermions by qubits.
Appendix~\ref{app:H2} illustrates our  methods using 
the hydrogen molecule as an example. 
Appendix~\ref{app:decoding} shows how to compute matrix elements
of simulator Hamiltonians constructed from LDPC codes. 

\section{Notations}
\label{sec:notations}

A  system of $M$ fermi  modes is described 
by the Fock space $\calF_M$  of dimension 
$2^M$ equipped with the standard basis 
$|x\rangle\equiv |x_1,\ldots,x_M\rangle$, where
 $x_\alpha=0,1$ is the occupation number 
of the mode $\alpha$  such that  $a_\alpha^\dag a_\alpha |x\rangle =x_\alpha |x\rangle$.
Our target Hilbert space is defined as the $N$-particle subspace of $\calF_M$,
\begin{equation}
\label{Nparticle}
\calH_{tgt}=\mathrm{span}{\left( |x\rangle \in \calF_M \, : \, |x|=N \right)}.
\end{equation}
Here $|x|$ denotes the Hamming weight of $x$. The simulator system
consists of $Q$ qubits, where $Q$ satisfies
\[
\dim{(\calH_{tgt})}={ M \choose N} \le 2^Q \le 2^M.
\]
The Hilbert space $\calH_{sim}=(\bC^2)^{\otimes Q}$ is equipped with the standard basis
$|s\rangle$, where $s\in \{0,1\}^Q$.
We shall reserve letters $s,t$ for qubit basis vectors
and letters $x,y$ for the Fock basis vectors. 
For any integer $K\ge 1$ let $[K]\equiv \{1,2,\ldots,K\}$.

Suppose  $\calO_{tgt}$ is a two-body or four-body  fermionic observable 
(hermitian operator)
listed below
\begin{equation}
\label{fewbody}
i^\epsilon(a_\alpha^\dag a_\beta \pm  a_\beta^\dag a_\alpha),
\quad i^{\epsilon}(a_\alpha^\dag a_\beta^\dag  a_\gamma a_\delta
\pm a_\delta^\dag a_\gamma^\dag  a_\beta a_\alpha),
\end{equation}
where $\epsilon=0,1$ is chosen to make the operator hermitian. 
We shall say  that 
a qubit observable $\calO_{sim}$ acting on $\calH_{sim}$ is a {\em simulator} of $\calO_{tgt}$ if 
\begin{equation}
\label{simulation1}
\calO_{sim} \calE = \calE \calO_{tgt}.
\end{equation}
A direct consequence of Eq.~(\ref{simulation1}) is that 
$\calO_{sim}$ preserves the codespace
and the restriction of $\calO_{sim}$ onto the codespace is unitarily equivalent
to $\calO_{tgt}$.  The action of $\calO_{sim}$ on the orthogonal complement
of the codespace may be arbitrary. 
Let us say that the simulator $\calO_{sim}$ is {\em $r$-sparse} if
it can be written as 
\begin{equation}
\label{sparse}
\calO_{sim}=\sum_{i=1}^r  D_i,
\end{equation}
where $D_i$ are hermitian operators such that each operator
$D_i$  is  diagonal in some tensor product basis of $Q$ qubits.
This basis may depend on $i$. 
The maximum sparsity $r$ of two-body and four-body simulators 
will be denoted $r_2$ and $r_4$ respectively.

\section{Sparse encodings for  small number of particles}
\label{sec:small}

Here we discuss encodings based on the first quantization method.
Assume for simplicity that the number of modes $M$ is a power of two, $M=2^m$.
Otherwise, round $M$ up to the nearest power of two. 
Given a fermi mode $\alpha \in \{1,\ldots, M\}$, let $\overline{\alpha}\in \{0,1\}^m$
be the binary representation of the integer $\alpha-1$.

The simulator system consists of $Q=mN$ qubits 
partitioned   into   $N$ consecutive registers
$Q_1,\ldots,Q_N$ of $m$ qubits each. 
For any $N$-tuple of modes $\alpha_1,\ldots,\alpha_N$ let
$|\overline{\alpha}_1,\overline{\alpha}_2, \cdots, \overline{\alpha}_N\rangle\in \calH_{sim}$
be a basis vector such that the register $Q_i$ is in the state $|\overline{\alpha}_i\rangle$.

Consider a basis vector $|x\rangle\in\calH_{tgt}$ and let
\[
1\le \alpha_1<\alpha_2<\ldots<\alpha_N\le M
\]
be the subset of $N$ modes
that are occupied in the state $|x\rangle$, that is,
$x_i=1$ iff $i\in\{\alpha_1,\ldots,\alpha_N\}$.
Define the encoding as
\begin{equation}
\label{small1}
\calE |x\rangle =\frac1{\sqrt{N!}}\sum_{\pi \in S_N} (-1)^{\pi} P_\pi
|\overline{\alpha}_1,\overline{\alpha}_2, \cdots, \overline{\alpha}_N\rangle,
\end{equation}
where $S_N$ is the group of permutations of $N$ objects,
$(-1)^\pi$ is the sign of a permutation $\pi$,
and $P_\pi$ is a unitary operator that applies a permutation
$\pi$ to the registers $Q_1,\ldots,Q_N$ such that 
\[
P_\pi |\overline{\alpha}_1,\overline{\alpha}_2, \cdots, \overline{\alpha}_N\rangle
=|\overline{\alpha}_{\pi(1)},\overline{\alpha}_{\pi(2)}, \cdots, \overline{\alpha}_{\pi(N)}\rangle.
\]
The righthand side of Eq.~(\ref{small1}) can be viewed as  the first-quantized version
of the state $|x\rangle$. The codespace $\mathrm{Im}(\calE)$ is 
spanned by  antisymmetric states $\psi\in \calH_{sim}$ such that 
\[
P_\pi |\psi\rangle = (-1)^\pi |\psi\rangle, \quad \mbox{for all $\pi\in S_N$}.
\]

We choose the simulator Hamiltonian as 
\begin{equation}
\label{small2}
H_{sim}=T+U+gH^{\perp},
\end{equation}
where $T+U$ is the first-quantized version of $H_{tgt}$,  namely
\begin{equation}
\label{T1}
T=\sum_{i=1}^N \sum_{\alpha,\beta=1}^M t_{\alpha\beta} |\alpha\rangle\langle \beta|_i
\end{equation}
\begin{equation}
\label{U1}
U=-\sum_{1\le i\ne j\le N}\;  \sum_{\alpha,\beta,\gamma,\delta=1}^M \; 
u_{\alpha\beta\gamma\delta} |\alpha,\beta \rangle\langle \gamma,\delta|_{i,j} 
\end{equation}
Here and below the subscripts $i,j$ indicate the registers $Q_i,Q_j$
acted upon by an operator. 
Finally, $H^{\perp}$ penalizes states orthogonal to
the codespace,
\begin{equation}
\label{Hperp}
H^{\perp}=\sum_{1\le i<j\le N}\; \frac12(I+(\leftrightarrow)_{i,j}).
\end{equation}
Here $(\leftrightarrow)_{i,j}$ is the SWAP operator that exchanges 
$Q_i$ and $Q_j$.  Note that $H^{\perp}$
has zero ground energy and its ground subspace coincides with the
codespace $\mathrm{Im}(\calE)$. The coefficient $g>0$ in Eq.~(\ref{small2})
will be chosen large enough so that the ground state of $H_{sim}$ belongs to the
codespace.

Note that $[T,P_\pi]=[U,P_\pi]=0$ for any permutation $\pi\in S_N$.
Thus  $H_{sim}$ preserves the codespace $\mathrm{Im}(\calE)$.
The standard correspondence between the first and the second quantized Hamiltonians
implies that the restriction of $H_{sim}$ onto the codespace is unitarily equivalent
to $H_{tgt}$, so that Eq.~(\ref{simulation}) is satisfied. Thus $H_{sim}$ is indeed a 
simulator of $H_{tgt}$ (for large enough $g$ to be chosen later).

Let us show that $H_{sim}$ is $r$-sparse, where 
\begin{equation}
\label{sparseA}
r=9^m \le M^{3.17}.
\end{equation}
Furthermore, 
$H_{sim}=\sum_{i=1}^r D_i$,
where each  term $D_i$ is diagonal in a tensor product of Pauli bases
\[
\calX\equiv \{ (|0\rangle \pm |1\rangle)/\sqrt{2}\}, \quad \calY\equiv \{  (|0\rangle \pm i|1\rangle)\sqrt{2}\},
\]
and $\calZ\equiv \{ |0\rangle, |1\rangle\}$.
Let 
\[
\mathrm{Pauli}(m)=\{\sigma_1,\sigma_2,\ldots,\sigma_{M^2} \}
\]
be the set of all $m$-qubit Pauli operators (ignoring the overall phase).
By definition, each operator $\sigma_a$ is a tensor product of
single-qubit Pauli operators $I,\sigma^x,\sigma^y,\sigma^z$.
Note that there are $4^m=M^2$ such operators.
We note that the SWAP operator on two qubits can be written as
\[
\frac12 \left( I\otimes I + \sigma^x \otimes \sigma^x + \sigma^y \otimes \sigma^y+\sigma^z \otimes \sigma^z\right).
\]
Since the SWAP operator $(\leftrightarrow)_{i,j}$ exchanging $m$-qubit registers
$Q_i$, $Q_j$ is a tensor product of $m$ two-qubit SWAPS, one gets
\begin{equation}
\label{swapPauli}
(\leftrightarrow)_{i,j} = M^{-1} \sum_{a=1}^{M^2} (\sigma_a\otimes \sigma_a)_{i,j}.
\end{equation}
Expanding each term in Eqs.~(\ref{T1},\ref{U1}) in the basis of Pauli operators and using
Eq.~(\ref{swapPauli}) to expand $H^{\perp}$ one gets
\begin{equation}
\label{small3}
H_{sim}=\sum_{1\le i<j\le N} \; \sum_{a,b=1}^{M^2} c_{a,b} (\sigma_a\otimes \sigma_b)_{i,j}
\end{equation}
for some real coefficients $c_{a,b}$.
We shall group Pauli operators that appear in Eq.~(\ref{small3}) into $r$ bins such that 
Pauli operators from the same bin are diagonal in the same tensor product basis. 
First consider a single register $Q_i$. Obviously, any Pauli operator acting on $Q_i$
is diagonal in a tensor product of the bases $\calX,\calY,\calZ$. 
Such tensor product bases can be labeled by letters in the alphabet
\begin{equation}
\label{A}
\calA\equiv \{\calX,\calY,\calZ\}^m.
\end{equation}
Recall that an  {\em orthogonal array}~\cite{Hedayat2012orthogonal} 
over an alphabet $\calA$
is a matrix $R$ of size $n\times k$ with entries from $\calA$
such that  any pair of columns of $R$ contains each two-letter word
in the alphabet $\calA$ the same number of times.
(More precisely, the above defines an orthogonal array with strength two). 
Orthogonal arrays have been previously used
for quantum simulations and dynamical decoupling~\cite{Wocjan2002,Kern2005,Rotteler2006}. 
We shall use a family of 
 orthogonal arrays based on the Galois field $GF(3^m)$
known as Rao-Hamming construction~\cite{Hedayat2012orthogonal}.
It gives an orthogonal $n\times k$ array  $R$ over an alphabet $\calA$ of size $3^m$
with $n=9^m$ and $k=3^m+1$.
Note that the equality $n=9^m$ is possible only if any pair of columns of $R$ contains each two-letter word
in the alphabet $\calA$ exactly one time. Also 
note that the number of particles $N$ obeys 
$N\le M=2^m<k=3^m+1$. We shall use only the first $N$ columns of $R$.
Let $f=1,\ldots,9^m$ be some row of $R$.
It defines a tensor product of Pauli bases 
\[
R_f\equiv R_{f,1}\otimes R_{f,2} \otimes \cdots \otimes R_{f,N}
\]
for the system of $Q$ qubits. 
By construction, each Pauli term $c_{a,b} (\sigma_a\otimes \sigma_b)_{i,j}$ that appears in  Eq.~(\ref{small3}) 
is diagonal in at least one basis $R_f$.
Thus we can choose a decomposition $H_{sim}=\sum_{f=1}^{9^m} D_f$
where $D_f$ is diagonal in the basis $R_f$.
This shows that any target Hamiltonian Eq.~(\ref{Htgt}) has a simulator with  sparsity 
$r=9^m=M^{2\log_2{(3)}}\approx M^{3.17}$.
Rounding $M$ up to the nearest power of two gives Eq.~(\ref{result1}).

The coefficient $g$ in Eq.~(\ref{small2}) must be  large enough so that
the ground state of $H_{sim}$ belongs to the codespace. 
This is always the case if $g\Delta^{\perp} > 2\|U+T\|$,
where $\Delta^{\perp}$ is the smallest non-zero eigenvalue of $H^\perp$,
see Eq.~(\ref{Hperp}).
Indeed, suppose $\psi$ is an  eigenvector of $H_{sim}$ orthogonal to the codespace.
Then the corresponding eigenvalue is 
\[
\langle \psi |H_{sim} |\psi\rangle \ge \langle \psi |U+T|\psi\rangle +g\Delta^\perp
>\|U+T\|.
\]
Such eigenvalue cannot be the ground
energy of $H_{tgt}$ since the latter is unitarily equivalent to a submatrix of $U+T$.
Thus the ground state of $H_{sim}$ must  belong to the codespace.
Recall that we assume the coefficients $t_{\alpha\beta},u_{\alpha\beta\gamma\delta}$ to
have magnitude at most one. This gives a conservative estimate
$\|U+T\|=O( N^2M^4)$.  
Below we show that 
\begin{equation}
\label{gap1}
\Delta^\perp=\frac{N}2 \quad \mbox{for all $N\le M$}.
\end{equation}
Thus it suffices to choose $g>4N^{-1}\|U+T\|=O(NM^4)$
and all terms $D_i$ in the simulator Hamiltonian have norm $poly(M)$. 
We do not expect the bound $g>O(NM^4)$ to be tight.
In practical implementation of variational quantum algorithms one can start from $g=0$ and gradually increase  $g$
until the best variational state $\psi$ satisfies $\langle \psi| H^\perp|\psi\rangle=0$.

Let us now prove Eq.~(\ref{gap1}). Recall that $\Delta^\perp$ is the smallest non-zero
eigenvalue of the Hamiltonain $H^\perp$ defined in Eq.~(\ref{Hperp}).
We shall use a  symmetry-based argument to compute  all eigenvalues of $H^\perp$.
Consider the permutation group $S_N$ and the unitary group $\calU(M)$.
The Hilbert space $\calH_{sim}\cong (\bC^M)^{\otimes N}$ defines a unitary representation
of these groups such that a permutation $\pi \in S_N$ acts on $\calH_{sim}$
as $P_\pi$ and a unitary matrix $V\in \calU(M)$ acts on $\calH_{sim}$ as $V^{\otimes N}$.
The standard result from the group representation theory known as
Schur duality~\cite{Goodman2000representations} gives a decomposition
\begin{equation}
\label{Schur}
(\bC^M)^{\otimes N}=\bigoplus_\lambda \calP_\lambda \otimes \calQ_\lambda,
\end{equation}
where the sum runs over all Young diagrams with $N$ boxes,
$\calP_\lambda$ is the irreducible representation (irrep) of the 
permutation group $S_N$ and $\calQ_\lambda$ is the irrep of the unitary group $\calU(M)$.
The operators $P_\pi$ and $V^{\otimes N}$ are block-diagonal
with respect to Schur decomposition Eq.~(\ref{Schur}).
Furthermore,  within each sector $\lambda$ the operator $P_\pi$ acts
non-trivially only on the subsystem $\calP_\lambda$, while
$V^{\otimes N}$ acts non-trivially only on subsystem  $\calQ_\lambda$.

Importantly, the Hamiltonian $H^\perp$ commutes with the action of both groups
$S_N$ and $\calU(M)$, that is,
\[
[H^\perp, P_\pi]=[H^\perp,V^{\otimes N}]=0
\]
for all $\pi\in S_N$ and for all $V\in \calU(M)$.
Since the groups $S_N$ and $\calU(M)$ generate 
the full operator algebra in each sector $\lambda$
in the decomposition Eq.~(\ref{Schur}),
we conclude that 
\begin{equation}
\label{Schur1}
H^\perp=\bigoplus_\lambda e_\lambda \Pi_\lambda,
\end{equation}
where $\Pi_\lambda$ is the projector onto the sector $\lambda$ in Eq.~(\ref{Schur})
and $e_\lambda$ are eigenvalues of $H^\perp$.
Thus we can compute $e_\lambda$ by picking an arbitrary state $\psi_\lambda$
from the sector $\lambda$ and computing  $e_\lambda=\langle \psi_\lambda |H^\perp|\psi_\lambda\rangle$.

\begin{figure}[h]
\centerline{\includegraphics[height=3.5cm]{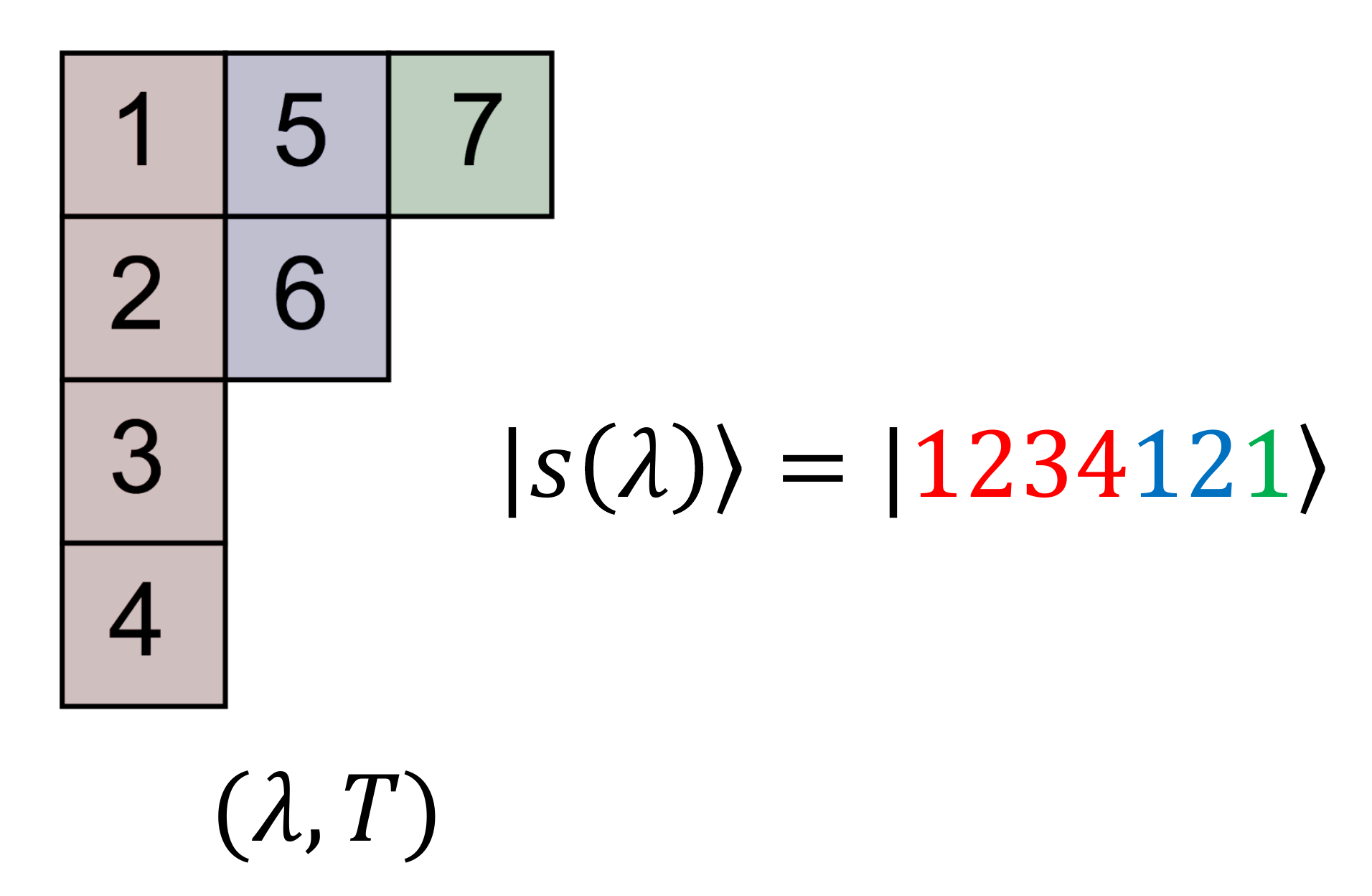}}
\caption{Example of a Young tableau $(\lambda,T)$ and the 
basis vector $|s(\lambda)\rangle$. Here  $N=7$ and $(\lambda_1,\lambda_2,\lambda_3)=(4,2,1)$.}
\label{fig:young}
\end{figure}

We shall consider a Young diagram $\lambda$ with $d$ columns as a partition of the integer $N$, that is,
\[
N=\lambda_1+\ldots +\lambda_d, \quad \lambda_1\ge \ldots\ge \lambda_d\ge 1.
\]
Namely, $\lambda_p$ is the number of boxes in the $p$-th column of $\lambda$.
For any integer $u$ define a state $|\phi(u)\rangle\in (\bC^M)^{\otimes u}$ such that 
\[
|\phi(u)\rangle = \frac1{\sqrt{u!}} \sum_{\pi\in S_u} (-1)^{\pi} |\pi(1),\pi(2),\ldots,\pi(u)\rangle.
\]
For example, $|\phi(1)\rangle=|1\rangle$, $|\phi(2)\rangle=(|1,2\rangle-|2,1\rangle)/\sqrt{2}$,
\[
\begin{array}{rccl}
|\phi(3)\rangle &=& \frac1{\sqrt{6}}  \left(  \right. & |1,2,3\rangle + |3,1,2\rangle + |2,3,1\rangle -  \\
&&  & \left. |2,1,3\rangle - |3,2,1\rangle - |1,3,2\rangle \right) \\
\end{array}
\]
etc.  We claim that a state 
\begin{equation}
\label{Schur2}
|\psi_\lambda\rangle \equiv |\phi(\lambda_1)\rangle\otimes \cdots \otimes  |\phi(\lambda_d)\rangle
\end{equation}
belongs to the sector $\lambda$ of the decomposition Eq.~(\ref{Schur}). 
Namely, such state can be obtained by applying a suitable Young symmetrizer~\cite{bacon2005quantum}
to a basis vector. Indeed, define a Young tableau $(\lambda,T)$ obtained by
filling columns of $\lambda$ one by one with consecutive integers
$1,\ldots,N$ starting from the first column, see
Figure~\ref{fig:young} for an example.
Let $S_{col}\subseteq S_N$ and $S_{row}\subseteq S_N$
be the subgroups that permute integers from the same column and from the same row 
of $(\lambda,T)$ respectively.
The Young symmetrizer corresponding to tableau $(\lambda,T)$  is defined as 
\begin{equation}
\label{symmetrizer}
\Pi_{\lambda,T}\sim  \left( \sum_{\pi \in S_{col}}\; (-1)^\pi P_\pi \right)
\cdot  \left( \sum_{\tau \in S_{row}}\;  P_\tau \right).
\end{equation}
It is well-known~\cite{Goodman2000representations,bacon2005quantum}
that $\Pi_{\lambda,T}$  is proportional to a (non-orthogonal) projector onto a subspace
of the sector $\lambda$ in the Schur decomposition.
In particular, $\Pi_{\lambda,T}$ maps any state
to some state that belongs to the sector $\lambda$. 
Let $s(\lambda)$  be a sequence of $N$ integers obtained by filling
columns of $\lambda$ one by one with consecutive integers 
such that the $j$-th column is filled with integers  $1,\ldots,\lambda_j$,
see Figure~\ref{fig:young} for an example.
Let $|s(\lambda)\rangle \in (\bC^M)^{\otimes N}$ be the basis
vector corresponding to $s(\lambda)$. 
We observe that the second factor in Eq.~(\ref{symmetrizer})
has trivial action on $|s(\lambda)\rangle$ since 
$P_\tau |s(\lambda)\rangle=|s(\lambda)\rangle$
for all $\tau\in S_{row}$.
It follows that 
\begin{equation}
\label{Schur3}
\Pi_{\lambda,T} |s(\lambda)\rangle  \sim |\psi_\lambda\rangle
\end{equation}
and thus $|\psi_\lambda\rangle$ indeed belongs to the sector 
$\lambda$ of the Schur decomposition.
One can easily check that $|\psi_\lambda\rangle$ is an
eigenvector of $H^\perp$ with the eigenvalue 
\begin{equation}
\label{Schur4}
e_\lambda=\frac12\left[ {N \choose 2} -\sum_{a=1}^d { \lambda_a \choose 2} + \sum_{1\le a<b\le d} 
\lambda_b \right].
\end{equation}
From Eq.~(\ref{Schur1}) one infers that any eigenvalue of $H^\perp$ must have a form
$e_\lambda$.  Note that $e_\lambda=0$  iff $\lambda$ is a single
column, that is, $d=1$, $\lambda_1=N$. One can check that the smallest non-zero
value of $e_\lambda$ is achieved when $\lambda$ has two columns with length $N-1$ and $1$,
that is, $d=2$, $\lambda_1=N-1$, and $\lambda_2=1$. In this case $e_\lambda=N/2$
which proves Eq.~(\ref{gap1}).

\section{Sparse encoding based on LDPC codes}
\label{sec:LDPC}

In this section we use the second quantization method and classical LDPC codes
to construct sparse encodings for the case when  the target system has
a constant filling fraction $\nu=N/M$.

Let  $A$ be a binary matrix with $Q$ rows and $M$ columns.
Given a binary vector $x$ of length $M$, 
we shall write $Ax$ for the matrix-vector multiplication modulo two, that is,
\begin{equation}
\label{Ax}
(Ax)_i=\sum_{\alpha=1}^M A_{i,\alpha} x_\alpha {\pmod 2}
\end{equation}
 We consider encodings $\calE\, : \, \calH_{tgt} \to \calH_{sim}$ defined by
\begin{equation}
\label{Edef}
\calE |x\rangle =|Ax\rangle
\end{equation}
where $x\in \{0,1\}^M$ and $|x|=N$.
Let us say that a matrix
$A$ is {\em $N$-injective} if it maps
distinct $M$-bit vectors $x$ with the Hamming weight $N$ to distinct $Q$-bit vectors $s=Ax$. 
It follows directly from the definitions that $\calE$ is an isometry
iff $A$ is $N$-injective. 
The $N$-injectivity condition is satisfied  if $A$ is chosen as a parity check matrix describing a binary linear code
of length $M$ with the minimum distance $2N+1$.   
Indeed, in this case all errors $x$ of weight up to $N$ must have different syndromes 
$s=Ax$ and thus a syndrome $s$ uniquely identifies a weight-$N$ error $x$.

 How sparse is the encoding defined in Eq.~(\ref{Edef})?
Let columns of $A$ be $A^1,\ldots,A^M\in \{0,1\}^Q$ and  let
\[
c(A)=\max_\alpha |A^\alpha|
\]
be the maximum column weight. We claim that
 fermionic observables defined in Eq.~(\ref{fewbody}) 
have simulators Eq.~(\ref{sparse})
with sparsity 
\begin{equation}
\label{r2r4}
r_2\le 2^{2c(A)-1} \quad \mbox{and} \quad r_4\le 2^{4c(A)-1}
\end{equation}
for two-body and four-body observables respectively. 
Furthermore, $\|D_i\|\le 1$ for all $i$.
Thus  the encoding defined by Eq.~(\ref{Edef}) is sparse whenever 
$A$ is a  column-sparse matrix.

First, let us introduce some notations. 
Let $A^{-1}s$ be a set of all weight-$N$ vectors $x$ satisfying $Ax=s$,
\begin{equation}
\label{invA}
A^{-1}s \equiv \{ x\in \{0,1\}^M\, : \, Ax=s \quad \mbox{and} \quad |x|=N\}.
\end{equation}
The set $A^{-1}s$ may be empty for some $s$. 
By definition,  $A$ is $N$-injective iff the set $A^{-1}s$ contains at most one element 
for any $s\in \{0,1\}^Q$. Below $e^\alpha=(0\ldots010\ldots0)$ denotes a string with a single non-zero
at the position $\alpha$.  We  use the notation  $\oplus$ for the bitwise XOR.

For concreteness, consider a pair of modes $\alpha<\beta$ and a target observable 
\begin{equation}
\label{tgt2body}
\calO_{tgt}=a_\alpha^\dag a_\beta+a_\beta^\dag a_\alpha.
\end{equation}
We have $\calO_{tgt}|x\rangle=0$ if $x_\alpha x_\beta =00,11$ and
\[
\calO_{tgt}|x\rangle = S_{\alpha\beta}(x)|x\oplus e^\alpha\oplus e^\beta\rangle
\]
if $x_\alpha x_\beta =01,10$
where $S_{\alpha\beta}(x)=\pm 1$ is the parity of all bits of $x$
located between $\alpha$ and $\beta$, that is,
\[
S_{\alpha\beta}(x)=\prod_{\gamma=\alpha+1}^{\beta-1} (-1)^{x_\gamma}.
\]
Since $A(x \oplus e^\alpha\oplus e^\beta)=Ax\oplus A^\alpha \oplus A^\beta$, we have
\begin{eqnarray}
\calE \calO_{tgt} |x\rangle &=&
S_{\alpha \beta}(x)  |Ax\oplus A^\alpha \oplus A^\beta \rangle \quad \mbox{if $x_\alpha x_\beta=01,10$}, \nonumber \\
\calE \calO_{tgt} |x\rangle&=& 0\quad \mbox{if $x_\alpha x_\beta=00,11$}.\label{trivial}
\end{eqnarray}
Let us say that a basis vector $s\in \{0,1\}^Q$ is $\alpha\beta$-flippable 
if $s=Ax$ for some weight-$N$ string $x$ such that $x_\alpha x_\beta=01$
or $x_\alpha x_\beta=10$. Note that  $A^{-1}s$ is a single string
whenever $s$ 
is $\alpha\beta$-flippable. Define an operator $\Gamma_{\alpha\beta}$
acting on $\calH_{sim}$ such that 
\begin{eqnarray}
\Gamma_{\alpha\beta} |s\rangle &=& S_{\alpha \beta}(A^{-1} s) |s\rangle 
\quad \mbox{if $s$ is $\alpha\beta$-flippable} \nonumber  \\
\Gamma_{\alpha\beta} |s\rangle&=&0  \quad \mbox{otherwise}. \label{G0}
\end{eqnarray}
Given a bit string $s$, let
$X(s)$ be the product of Pauli $\sigma^x$ operators over all qubits $i$
such that $s_i=1$. 
We claim that the observable $\calO_{tgt}$ has a simulator
\begin{equation}
\label{sim2body}
\calO_{sim}=X(A^\alpha \oplus A^\beta) \Gamma_{\alpha\beta} = \Gamma_{\alpha\beta}X(A^\alpha \oplus A^\beta).
\end{equation}
First let us check that $X(A^\alpha \oplus A^\beta)$ commutes with $\Gamma_{\alpha\beta}$.
Suppose $s$ is $\alpha\beta$-flippable and let $t=s\oplus A^\alpha\oplus A^\beta$.
By assumption, 
$s=Ax$ for some $x$ such that $|x|=N$
and, say, $x_\alpha x_\beta=01$. It follows that $y\equiv x\oplus e^\alpha\oplus e^\beta$
has weight $N$ and $Ay=t$.
Furthermore, $y_\alpha y_\beta=10$.
Thus  $t$ is $\alpha\beta$-flippable and $A^{-1}t=y$.
Since $S_{\alpha\beta}(x)=S_{\alpha\beta}(y)$, we have shown that 
$S_{\alpha\beta}(A^{-1}s)=S_{\alpha\beta}(A^{-1}t)$ and thus 
\[
X(A^\alpha \oplus A^\beta) \Gamma_{\alpha\beta} |s\rangle=S_{\alpha\beta}(x)|t\rangle=
\Gamma_{\alpha\beta}X(A^\alpha \oplus A^\beta)|s\rangle.
\]
If $s$ is not $\alpha\beta$-flippable then so is $t$, so that 
\[
X(A^\alpha \oplus A^\beta) \Gamma_{\alpha\beta} |s\rangle=
\Gamma_{\alpha\beta}X(A^\alpha \oplus A^\beta)|s\rangle=0.
\]
We have shown that
$X(A^\alpha \oplus A^\beta)$ commutes with $\Gamma_{\alpha\beta}$.

Next, let us check the simulation condition Eq.~(\ref{simulation}).
Suppose $x$ has weight $N$ and let $s=Ax$.
Using the first equality in Eq.~(\ref{sim2body}) one infers
that 
\[
\calO_{sim}\calE|x\rangle=X(A^\alpha\oplus A^\beta)\Gamma_{\alpha\beta}|s\rangle=
S_{\alpha\beta}(x) |s\oplus A^\alpha\oplus A^\beta \rangle
\]
if $x_\alpha x_\beta=01,10$ and $\calO_{sim}\calE |x\rangle=0$ 
otherwise. Comparing this and Eq.~(\ref{trivial}) shows that
$\calO_{sim} \calE=\calE \calO_{tgt}$. 

Let us show that $\calO_{sim}$ is $r$-sparse 
with $r\le 2^{2c(A)-1}$.   By construction, $\Gamma_{\alpha\beta}$
is diagonal in the standard basis, that is,
\begin{equation}
\label{G1}
\Gamma_{\alpha\beta}=\sum_{s\in \{0,1\}^M} \; g(s) |s\rangle\langle s|, \qquad g(s)=0,\pm 1.
\end{equation}
Let $k\equiv |A^\alpha \oplus A^\beta|$. To simplify  notations, let us reorder the qubits
such that $X(A^\alpha \oplus A^\beta)=X^{\otimes k}$ acts on the first $k$ qubits. 
Decompose $s=(u,s')$, where 
$u\in \{0,1\}^k$ and $s'\in \{0,1\}^{Q-k}$.
Define a function $h(t,s')$ obtained from $g(u,s')$ by applying the Walsh-Hadamard
transform with respect to the first argument:
\begin{equation}
\label{WH}
h(t,s')\equiv 2^{-k} \sum_{u\in \{0,1\}^k}\; (-1)^{t\cdot u} g(u,s').
\end{equation}
Here $t\cdot u\equiv \sum_{i=1}^k t_i u_i$. Substituting the identity
\[
|s\rangle\langle s|\equiv |u,s'\rangle\langle u,s'| = 2^{-k} \sum_{t\in \{0,1\}^k} \; (-1)^{t\cdot u} Z(t)\otimes |s'\rangle\langle s'|
\]
into Eq.~(\ref{G1}) gives
\begin{equation}
\label{G2}
\Gamma_{\alpha\beta}=\sum_{t\in \{0,1\}^k} \; \sum_{s'\in \{0,1\}^{Q-k}}\; h(t,s') Z(t)\otimes |s'\rangle\langle s'|.
\end{equation}
For each $t\in \{0,1\}^k$ define an operator 
\begin{equation}
\label{G3}
\Gamma_{\alpha\beta}(t)=\sum_{s\in \{0,1\}^{Q-k}}\; h(t,s)  |s\rangle\langle s|.
\end{equation}
acting on the last $Q-k$ qubits. Then
\begin{equation}
\label{G4}
\Gamma_{\alpha\beta}=\sum_{t\in \{0,1\}^k}\; Z(t) \otimes \Gamma_{\alpha\beta}(t).
\end{equation}
As was shown above, $X(A^\alpha\oplus A^\beta)=X^{\otimes k}$ commutes with $\Gamma_{\alpha\beta}$.
Since $X^{\otimes k}$ commutes (anti-commutes) with $Z(t)$ for even (odd) $t$,
we infer from Eq.~(\ref{G4}) that 
$\Gamma_{\alpha\beta}(t)=0$ whenever $t$ has odd weight. 
Combining Eqs.~(\ref{sim2body},\ref{G4}) we arrive at
\begin{equation}
\label{G5}
\calO_{sim} =\sum_{\substack{t\in \{0,1\}^k\\ |t| \; \mathrm{even} \\}}\; \;  X^{\otimes k} Z(t) \otimes \Gamma_{\alpha\beta}(t),
\end{equation} 
This gives  a $2^{k-1}$-sparse decomposition of $\calO_{sim}$
as defined in Eq.~(\ref{sparse}) 
where 
\begin{equation}
\label{DiLDPC}
D_i\equiv X^{\otimes k} Z(t) \otimes \Gamma_{\alpha\beta}(t).
\end{equation}
This operator can be made diagonal in the standard basis by applying
a Clifford operator exchanging Pauli $Y$ and $Z$ for all qubits $i=1,\ldots,k$
such that $t_i=1$. It remains to note that 
\[
k=|A^\alpha\oplus A^\beta|\le |A^\alpha|+|A^\beta|\le 2c(A).
\]
Thus the simulator Eq.~(\ref{G5}) has sparsity $2^{2c(A)-1}$.
Furthermore, all  matrix elements of $D_i$ are contained
in the interval $[-1,1]$, see Eqs.~(\ref{G1},\ref{WH},\ref{G3},\ref{DiLDPC}).
We omit the derivation of simulators for other  observables defined in Eq.~(\ref{fewbody})
since it follows  exactly the same steps as above.

Consider a target Hamiltonian $H_{tgt}$ defined in  Eq.~(\ref{Htgt}). 
Decompose $H_{tgt}$ as a linear combination of two-body and four-body observables
$\calO_{tgt}$ 
defined in Eq.~(\ref{fewbody}). 
Replacing each observable $\calO_{tgt}$
by a qubit simulator $\calO_{sim}$ constructed above  gives a simulator Hamiltonian
\begin{equation}
\label{Hsim1}
H_{sim}=g\calE\calE^\dag + \sum_{i=1}^r D_i.
\end{equation}
Here we combined the terms $D_i$ from each simulator $\calO_{sim}$ into a single sum.
The term $g\calE\calE^\dag$ penalizes states orthogonal to the codespace.
Note that $r$ is upper bounded by a constant $O(1)$ times the number of non-zero
coefficients $t_{\alpha\beta},u_{\alpha\beta\gamma\delta}$ in the target Hamiltonian.
Since $\|D_i\|\le 1$, we can guarantee that the ground state of $H_{sim}$
belongs to the codespace provided that $g\sim M^4$. 

Let us choose $A$ as a parity check matrix of a binary linear code
that encodes $K$ bits
into $M$ bits with the minimum distance $2N+1<M$.
As was argued above, such matrix $A$ is $N$-injective. 
It is known that certain families of codes described by sparse parity check matrices
can approach the  Gilbert-Varshamov bound~\cite{Varshamov1957,MacWilliamsBook}, namely,
\begin{equation}
\label{GVbound}
K= M(1-h(2N/M)-\epsilon),
\end{equation}
where $h(x)=-x\log_2{(x)} -(1-x)\log_2{(1-x)}$  is the binary Shannon entropy function
and $\epsilon>0$ can be made arbitrarily small by choosing large enough $c(A)$.
This claim follows from the existence of good LDPC codes~\cite{Gallager1962journal},
see for instance Theorem~A.3 of~\cite{Gallager1962}.
We can assume wlog that all rows of $A$ are linearly independent
in which case $A$ has $Q=M-K$ rows. 
Then a family of good LDPC codes as above gives a family of sparse
encodings with the filling fraction $\nu=N/M$ and 
the qubit-per-mode ratio $\eta=Q/M=1-K/M$ that satisfy
$\eta=h(2\nu)+\epsilon$, as claimed in Eq.~(\ref{result1}).
Unfortunately, the constant $c(A)$ grows quickly as  $\epsilon$
approaches $0$, see~\cite{Gallager1962}.
Since the sparsity of  simulators constructed  for few-body fermionic
observables 
is exponential in $c(A)$, see Eq.~(\ref{r2r4}), encodings
based on good LDPC codes are not quite practical.
We show how  to overcome this problem using ``bad" LDPC codes
in Sections~\ref{sec:improved},\ref{sec:graph}.

Next let us  discuss how to compute 
matrix elements of the simulator Hamiltonian Eq.~(\ref{Hsim1}).
Note that all steps in the definition of $H_{sim}$ are 
computationally efficient except for inverting the action of $A$,
that is, computing the set $A^{-1} s$ defined in Eq.~(\ref{invA}).
Define a function 
\begin{equation}
\label{xmin}
x_{min}(s)=\arg \; \min_{\substack{ x\in \{0,1\}^M \\  Ax=s \\}} \; |x|.
\end{equation}
It  returns an error $x\in \{0,1\}^M$ of minimum  weight consistent with a given 
syndrome $s\in \{0,1\}^Q$. 
Suppose $A$ is a parity check matrix of a linear code 
with the minimum distance $2N+1$. It follows easily from the definitions
that $A^{-1}s=\{x_{min}(s)\}$  if $x_{min}(s)$ has weight $N$
and $A^{-1}s=\emptyset$ otherwise. 
Thus it suffices to give an efficient algorithm for computing $x_{min}(s)$. 
The latter is known as a {\em minimum weight decoding} problem. 
Although in general this problem  is NP-hard~\cite{Berlekamp1978}, 
there are special classes of LDPC codes that admit a linear time 
decoder~\cite{sipser1996expander,zemor2001expander}.
These codes have a non-zero encoding rate and relative distance, but 
they are not good in the sense of Eq.~(\ref{GVbound}).
In Section~\ref{sec:graph} we discuss a special class of LDPC codes
based on high-girth bipartite graphs that can be decoded in time $O(M^3)$.
Appendix~\ref{app:decoding} gives a simple algorithm
that computes the set $A^{-1}s$ for any $N$-injective matrix. 
Although this algorithm is not efficient asymptotically, 
it can be implemented for small system sizes $M\le 50$.

\section{Improving the sparsity}
\label{sec:improved}

Here we show how to improve the sparsity bounds in Eq.~(\ref{r2r4})
if the parity check matrix $A$ has a certain additional structure. At this point we shall exploit the fact that simulators
only need to reproduce the action of  target observables within the codespace
and can act arbitrarily on the orthogonal complement to the codespace.

Let $A$ be a binary matrix of size $Q\times M$
with columns $A^1,\ldots,A^M$. We shall say that $A$
is  {\em bipartite} if  the set of rows $[Q]\equiv \{1,\ldots,Q\}$ can be partitioned into two disjoint subsets,
$[Q]=L\cup R$, such that each column $A^\alpha$
intersects both $L$ and $R$ on odd number of rows,
\begin{equation}
\label{bipartite}
|A^\alpha \cap L | {\pmod 2} = |A^\alpha \cap R| {\pmod 2} =1
\end{equation}
for all $1\le \alpha\le M$.
We claim that  the encoding Eq.~(\ref{Edef}) based on a
bipartite matrix $A$ has sparsity parameters
\begin{equation}
\label{improved1}
r_2\le 2^{2c(A)-3} \quad \mbox{and} \quad r_4\le 2^{4c(A)-3}.
\end{equation}
Indeed, consider any weight-$N$ string $x$ and let $s=Ax$ be its syndrome. 
Let $s(L),s(R)\in \{0,1\}$ be the parity of $s$ restricted to $L$ and $R$,
\[
s(L)\equiv \sum_{i\in L} s_i {\pmod 2}\quad \mbox{and} \;
\quad
s(R)\equiv \sum_{i\in R} s_i {\pmod 2}.
\]
From Eq.~(\ref{bipartite}) one infers that flipping any bit of $x$
flips the values of $s(L)$ and $s(R)$. Therefore
\begin{equation}
\label{improved2}
s(L)=s(R)=(-1)^N
\end{equation}
are constants independent of $x$. 
Thus the codespace $\mathrm{Im}(\calE)$ is stabilized
by the products of Pauli $Z$ operators over $L$ and $R$,
\begin{equation}
\label{improved3}
Z(L)|s\rangle = Z(R)|s\rangle =(-1)^N |s\rangle \quad \mbox{for all $|s\rangle\in \mathrm{Im}(\calE)$}.
\end{equation}

Below we use notations and terminology of Section~\ref{sec:LDPC}.
Consider a two-body fermionic observable $\calO_{tgt}$ 
acting on a pair of modes $\alpha,\beta$, see  Eq.~(\ref{tgt2body}).
It has a $2^{k-1}$-sparse simulator defined by Eq.~(\ref{G5}) where
$k=|A^\alpha\oplus A^\beta|$. 
If $A^\alpha \cap A^\beta\ne \emptyset$ then $k\le |A^\alpha|+|A^\beta| -2 \le 2c(A)-2$
and thus the simulator has sparsity $2^{2c(A)-3}$, as claimed. 
From now on we assume that $A^\alpha \cap A^\beta= \emptyset$.
The assumption Eq.~(\ref{bipartite}) implies that $X(A^\alpha)$
anti-commutes with $Z(L)$ and $Z(R)$ for all $\alpha$. Therefore
$X(A^\alpha\oplus A^\beta)=X(A^\alpha) X(A^\beta)$ commutes
with $Z(L)$ and $Z(R)$. We shall modify the simulator $\calO_{sim}$ defined
in Eq.~(\ref{G5}) by multiplying some terms in Eq.~(\ref{G5})  by 
$(-1)^N Z(L)$, or $(-1)^N Z(R)$, or $Z(L)Z(R)$. 
As was argued above, these operators commute with each term in Eq.~(\ref{G5})
and have trivial action on the codespace due to Eq.~(\ref{improved3}). 
Thus we the modified simulator $\calO_{sim}'$ has exactly the same
action on the codespace as $\calO_{sim}$, that is, $\calO_{sim}'$
is a simulator of $\calO_{tgt}$. 

Fix some pair of qubits $i\in A^\alpha \cap L$ and $j\in A^\alpha \cap R$.
Multiply each term in Eq.~(\ref{G5}) with $t_i=1$ and $t_j=0$ by $(-1)^N Z(L)$.
Multiply each term in Eq.~(\ref{G5}) with $t_i=0$ and $t_j=1$ by $(-1)^N Z(R)$.
Multiply each term in Eq.~(\ref{G5}) with $t_i=1$ and $t_j=1$ by $Z(L)Z(R)$.
This cancels the action of Pauli $Z(t)$ in Eq.~(\ref{G5}) on the chosen pair of qubits $i,j$.
Thus we can write
\begin{equation}
\label{improved4}
\calO_{sim}' =\sum_{\substack{t\in \{0,1\}^k \\ |t| \;  \mathrm{even} \\ t_i=t_j=0\\}}\; \;  X^{\otimes k} Z(t) \otimes 
\Gamma_{\alpha\beta}'(t)
\end{equation}
for some new operators $\Gamma_{\alpha\beta}'(t)$ diagonal in the standard basis. 
Thus $\calO_{sim}'$ has sparsity  $2^{k-3}\le 2^{2c(A)-3}$ as claimed
in Eq.~(\ref{improved1}). We omit the derivation for other  observables defined in Eq.~(\ref{fewbody})
since it follows the same steps as above.

\section{Graph-based encodings}
\label{sec:graph}

Suppose $G$ is a bipartite graph with $Q$ vertices and $M$ edges.
We assume that vertices of $G$ are partitioned into two disjoint subsets $L,R$
such that only edges between $L$ and $R$ are allowed.
Let $A$ be the incidence matrix of $G$. By definition, 
$A$ has $Q$ rows, $M$ columns, and $A_{i,\alpha}=1$ if a vertex $i$
is an endpoint of an edge $\alpha$. Otherwise $A_{i,\alpha}=0$.
Consider the encoding $\calE|x\rangle=|Ax\rangle$.
Since $c(A)=2$ and $A$ is bipartite, few-body fermionic observables
have  simulators with sparsity $r_2=2$ and $r_4=32$, see 
Eq.~(\ref{improved1}).

Suppose the number of qubits $Q$ and the number of particles $N$ are fixed.
What is the maximum value of $M$ that can be achieved using encodings
based on bipartite graphs? First let us rephrase the $N$-injectivity condition 
in terms of the {\em girth} of the graph $G$. Recall that
a graph $G$ has girth $g$ if any closed loop in $G$ has at least $g$
edges. 

We claim that the matrix $A$ is $N$-injective iff the graph $G$ 
has girth $g\ge 2N+2$. Indeed, assume that $A$ is not $N$-injective. 
Then $Ax=Ay$ for some pair of weight-$N$ strings $x\ne y$. 
Let $z=x\oplus y$ so that $Az=0$ and $|z|\le 2N$. We can consider $z$ as a subset of edges in  $G$. 
From $Az=0$ one infers that $z$ is a cycle, that is, each vertex  has
even number of incident edges from $z$. 
However, each cycle contains at least one closed loop. If $z'\subseteq z$ is such a loop
then $|z'|\le |z|\le 2N$, that is, $g\le 2N$.
Conversely, assume that $g\le 2N$.
Let $z$ be any loop of length at most $2N$. Note that $z$ must have even length
since $G$ is bipartite. Choose any partition $z=x\oplus y$ such that $|x|=|y|=|z|/2$
and $x\cap y=\emptyset$. 
Then $Ax=Ay$. Choose any subset of edges $u$ such that 
$|u|=N-|z|/2$ and such that $x,y,u$ are pairwise disjoint. This is always possible
since 
\[
M-|x|-|y|=M-|z|\ge M-2N\ge 0.
\]
Let  $x'=x\oplus u$ and $y'=y\oplus u$. Then $x'\ne y'$,  $Ax'=Ay'$ and $|x'|=|y'|=N$, that is,
$A$ is not $N$-injective. This proves the claim. 

The above shows that maximizing $M$ for fixed $N$ and $Q$  is equivalent
to finding the largest  bipartite  graph with a fixed number
of vertices $Q$ and a girth $g\ge 2N+2$. 
This  problem has a long history in the graph theory, see for instance~\cite{Hoory2002size}
and the references therein. In particular, 
nearly maximal  bipartite graphs with a given girth can be constructed
by greedy algorithms~\cite{Bayati2009,Osthus2001}.
Such algorithms start from an empty graph and 
sequentially  add random edges drawn from a suitable  (time dependent) probability distribution.
The process terminates once  there is no edge that  can be added
without reducing the girth below the specified value, 
see Refs.~\cite{Bayati2009,Osthus2001}
for details. The data shown on Figure~\ref{figure:1} was generated
using the greedy algorithm of Ref.~\cite{Osthus2001} with $10^3$ trials for
each pair $Q,N$ and selecting the maximum graph with girth at least $2N+2$.
The number of edges $M$ in the maximum graph gives a lower bound on the
number of fermi modes that can be simulated for a given pair $Q,N$, see  Figure~\ref{figure:1}.

\begin{figure}[h]
\centerline{\includegraphics[height=3.5cm]{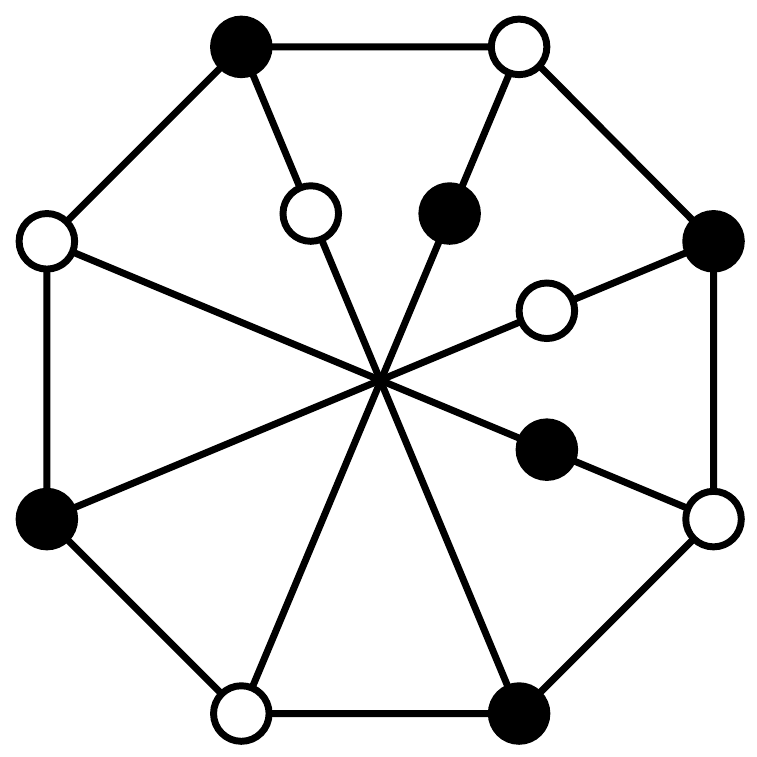}}
\caption{Example of a girth-$6$ bipartite graph with $12$ vertices and $16$ edges. 
Each  empty or filled circle is a vertex.}
\label{fig:example}
\end{figure}

As a simple example consider a girth-$6$ bipartite  graph shown on Figure~\ref{fig:example}.
It has $12$ vertices and $16$ edges. 
This graph encodes a system of $M=16$ fermi modes with $N=2$ particles
into a system of $Q=12$ qubits.
One can generalize this example as follows. 
Start from a cycle of even length $L$ such that $L\ge 2N+2$ and connect each pair of 
vertices $j$ and $j+L/2 {\pmod L}$ by a chord. 
Each chord contains $N-1$ vertices and $N$ edges. 
This defines a bipartite graph with girth $g=2N+2$ similar to the
one shown on Figure~\ref{fig:example}.
The graph has $M=L+NL/2$ edges and $Q=L+(N-1)L/2$ vertices.
Condition $L\ge 2N+2$ is equivalent to $(1+N)(2+N)\le M$. Thus 
\[
Q=M-\frac{M}{2+N} \quad \mbox{for $N\le M^{1/2}-O(1)$}.
\]
This encoding can eliminate approximately $M/N$ qubits.
We observe that some entries in the table of Figure~\ref{figure:1}
can be obtained using the above construction. For example,
the encoding with $Q=20$, $N=3$, $M=25$ correspond
to the graph of Figure~\ref{fig:example} where the cycle has length $L=10$
and each chord contains $N=3$ edges. Such graph has girth $g=8$
and $M=25$ edges.

In contrast to encodings based on general LDPC codes,  
graph-based encodings give simulator Hamiltonians with efficiently computable matrix elements. 
Indeed,  suppose $x$ is a minimum weight solution of the equation $Ax=s$, see Eq.~(\ref{xmin}).
If $A$ is an incidence matrix of a graph, one can view $x$ and $s$ as subsets of edges and vertices
respectively. Obviously, $x$ is minimal if it consists of edge disjoint paths connecting pairs
of vertices in $s$. Moreover, $x$ defines a perfect matching on the set $s$ such that
each matched pair of vertices in $s$ is connected by a shortest path.
Thus computing a minimum weight solution of $Ax=s$
is equivalent to (a) computing a shortest path between each pair of vertices in $s$
and (b) finding a minimum weight perfect matching of vertices of $s$.
These steps can be done in time $O(M^3)$ using the Dijkstra's algorithm
to compute the shortest paths and Edmonds blossom algorithm to find the
minimum weight perfect matching, see~\cite{schrijver2002combinatorial} for more details.

\section{Dicrete $\bZ_2$ symmetries -- particle and spin conservation}
\label{sec:parity}

In the next two sections we discuss encodings based on the Jordan-Wigner transformation~\cite{Wigner1928}
and its recent generalizations~\cite{BK2002,seeley2012bravyi}.
Such encodings are well suited for the removal of qubits in the presence of discrete $\bZ_2$ symmetries
such as those describing the fermionic parity conservation. Removal of qubits for the $\textrm{H}_2$ molecular 
Hamiltonian and a two-site Hubbard model were considered in~\cite{Moll2015}. 
We generalize the approach and consider a system of $M$ fermi modes and assume that our target Hilbert
space is the full Fock space $\calF_M$. 
The simulator system consist of $M$ qubits. 
We consider encodings $\calE\, : \, \calF_M\to (\bC^2)^{\otimes M}$ 
such that
\begin{equation}
\label{JW1}
\calE|x\rangle=|Ax\rangle \quad \mbox{for all} \quad x\in \{0,1\}^M,
\end{equation}
where $A$ is some $M\times M$ invertible binary matrix
and $Ax$ stands for the matrix-vector multiplication modulo two,  cf. Eqs.~(\ref{Ax},\ref{Edef}).
The standard Jordan-Winger transformation is obtained by choosing $A$ as the identity matrix, 
$A=I$. A binary tree and the parity  encodings introduced in Refs.~\cite{BK2002,seeley2012bravyi}
can be viewed as generalizations of the Jordan-Wigner transformation. 
A binary tree encoding is defined for $M$ being a power of two, $M=2^m$.
If $M\neq 2^m$, one refers to the definition for a number of modes $2^m>M$, using only the correspondences 
for the first $M$ modes. The binary tree encoding is obtained by choosing
$A\equiv A_m$, where a sequence of matrices $A_0,A_1,\ldots,A_m$ is defined
recursively~\cite{seeley2012bravyi} as
\begin{equation}
\label{Abk}
A_0=1, \quad A_1=\left[ \begin{array}{cc} 1 & 0 \\ 1 & 1\\ \end{array} \right],
\quad A_{k}=\left[ \begin{array}{cc}  A_{k-1} &  \boldsymbol{0}  \\
 B_{k-1} & A_{k-1}   \\ \end{array} \right],
\end{equation}
where $\boldsymbol{0}$ is the all-zeros matrix and 
 $B_k$ is a matrix of size $2^k\times 2^k$ that has the last row filled by ones
and all remaining rows filled by zeros. 
For example, choosing $M=4$ one gets 
\[
A=\left[ \begin{array}{cccc}
1 & 0 & 0 & 0 \\
1 & 1 & 0 & 0 \\
0 & 0 & 1 & 0 \\
1 & 1 & 1 & 1\\
\end{array}
\right].
\]
Finally, the parity encoding~\cite{seeley2012bravyi}
is obtained by choosing  $A$ as a lower-triangular
$M\times M$ matrix,
\begin{equation}
\label{Aparity}
A_{i,j}=\left\{ \begin{array}{rcl}
1 & \mbox{if} & i\ge j \\
0 && \mbox{otherwise} \\
\end{array}\right.
\end{equation}

For example, choosing $M=4$ one gets
\[
A=\left[ \begin{array}{cccc}
1 & 0 & 0 & 0 \\
1 & 1 & 0 & 0 \\
1 & 1 & 1 & 0 \\
1 & 1 & 1 & 1\\
\end{array}
\right].
\]
The main advantage of the binary tree encoding is that any few-body fermionic 
observable $\calO_{tgt}$ defined in Eq.~(\ref{fewbody}) has a qubit simulator
$\calO_{sim}=\calE \calO_{tgt} \calE^\dag$ such that $\calO_{sim}$ is a  Pauli-like operator
acting non-trivially only on $O(\log{M})$ qubits~\cite{BK2002}.
In contrast, the standard  Jordan-Wigner and the parity encodings can map
a few-body fermionic observable to a Pauli-like operator acting on all $M$ qubits,
see Ref.~\cite{seeley2012bravyi} and Appendix~\ref{app:StandardMaps} for more details. 

Consider a target Hamiltonian Eq.~(\ref{Htgt}) that describes  a molecule with $M$ spin-orbitals.
Accordingly,  each fermi mode $\alpha$ is a pair $\alpha=(i,\omega)$, where $i=1,\ldots,M/2$ is a spatial orbital and $\omega\in \{\uparrow,\downarrow\}$ is
the spin orientation.
It is well-known that
molecular Hamiltonians based on the non-relativistic
Schr\"odinger equation conserve the number of particles with a fixed spin orientation~\cite{szabo1989modern}.
Let us order the $M$ modes such that the first (the last) $M/2$ modes describe orbitals with spin up (spin down). Then
\begin{equation}
\label{sym2}
[H_{tgt},\hat{N}_\uparrow]=[H_{tgt},\hat{N}_\downarrow]=0,
\end{equation}
where
\be
\label{order}
\hat{N}_\uparrow=\sum_{\alpha=1}^{M/2} a_\alpha^\dag a_\alpha
\quad \mbox{and} \quad
\hat{N}_\downarrow=\sum_{\alpha=M/2+1}^{M} a_\alpha^\dag a_\alpha
\ee
are particle number operators for spin-up and spin-down modes.
We claim that the symmetry Eq.~(\ref{sym2}) can be exploited to remove two
qubits from the simulator Hamiltonian obtained via the binary tree encoding.
For a special case when $M=4$ and $H_{tgt}$ describes the hydrogen molecule
see Ref.~\cite{o2015scalable}.
Indeed, suppose $M=2^m$ and let $A=A_m$ be the matrix defined by Eq.~(\ref{Abk}).
We note that the $M$-th row of $A$ has a form $1^M$ (the all-ones string).
Furthermore, the row $M/2$ has a form $1^{M/2}0^{M/2}$.
 It follows that 
\begin{equation}
\label{Ax1}
(Ax)_M =\sum_{\alpha=1}^M x_\alpha  {\pmod 2}
\end{equation}
and
\begin{equation}
\label{Ax2}
(Ax)_{M/2}=\sum_{\alpha=1}^{M/2} x_\alpha  {\pmod 2}.
\end{equation}
Comparing Eqs.~(\ref{order},\ref{Ax1},\ref{Ax2}) one concludes that 
\begin{equation}
\label{Qsym}
\calE (-1)^{\hat{N}_\uparrow} \calE^\dag = \sigma^z_{M/2}
\quad \mbox{and} \quad
\calE (-1)^{\hat{N}_\uparrow+ \hat{N}_\downarrow} \calE^\dag =\sigma^z_M.
\end{equation}
Here and below $\sigma^x_j,\sigma^y_j,\sigma^z_j$ denote  Pauli 
operators acting on the $j$-th qubit. 
From Eqs.~(\ref{sym2},\ref{Qsym}) one infers that the simulator Hamiltonian
$H_{sim}=\calE H_{tgt} \calE^\dag$ commutes with $\sigma^z_M$ and $\sigma^z_{M/2}$. 
This means that  eigenvectors  of $H_{sim}$ can be chosen as eigenvectors of
$\sigma^z_M$ and $\sigma^z_{M/2}$. Accordingly, when looking for the ground energy
of $H_{sim}$ in a variational approach one can replace  Pauli operators
$\sigma^z_M$ and $\sigma^z_{M/2}$ by their eigenvalues $\pm 1$ and remove 
qubits $M$ and $M/2$ from the simulator system. 
Importantly, this removal does not affect locality properties of the binary tree encoding,
that is, the final Hamiltonian $H_{sim}$ acting on $M-2$ qubits is a linear combination
of Pauli-like operators with weight $O(\log{M})$. 
We note that exactly the same arguments as above can be applied to
remove two qubits from the simulator Hamiltonian obtained via the parity encoding Eq.~(\ref{Aparity})
since the matrices $A$ defined in Eqs.~(\ref{Abk},\ref{Aparity}) have the same
rows $M$ and $M/2$.

\section{Multiple $\bZ_2$ symmetries}
\label{sec:multiple}

The aforementioned qubit removal process  can be easily generalized
to Hamiltonians with multiple $\bZ_2$ symmetries. Such symmetries are usually  present in 
Hamiltonians describing molecules with geometric symmetries such as rotations or reflections,
see Table~\ref{tablemol} below.
 Let $H_{sim}$ be a qubit simulator Hamiltonian
obtained by applying some version of the Jordan-Wigner encoding Eq.~(\ref{JW1})
from the previous section. 
We can write $H_{sim}$ as a linear combination of $M$-qubit  Pauli operators 
\begin{equation}
\label{Hpauli}
H_{sim} =\sum_{j=1}^r h_j  \boldsymbol{\sigma}_j,
\end{equation}
Here $h_j$ are some real coefficients and $\boldsymbol{\sigma}_j$ are
$M$-qubit Pauli operators, that is, 
$M$-fold tensor product of  single-qubit Pauli operators $I,\sigma^x,\sigma^y,\sigma^z$. 
Let 
\[
\cP_M=\pm\{I,\sigma^x,\sigma^y,\sigma^z \}^{\otimes M}
\]
be the set of all $M$-qubit  Pauli operators.  Recall that the $M$-qubit Clifford group $\calC_M$ is defined as a set of unitary operators $U$
such that $U\boldsymbol{\sigma} U^\dag \in \cP_M$ for all $\boldsymbol{\sigma}\in \cP_M$.

Suppose $\calS\subseteq \cP_M$ is an abelian group.
We shall say that $\calS$ is a symmetry group of $H_{sim}$ if
any element of $\calS$ commutes with each Pauli term in $H_{sim}$
and $-I\notin \calS$. It is a well-known fact from the theory of stabilizer codes~\cite{gottesman1997stabilizer} 
that any abelian group $\calS\subseteq \calP_M$ such that $-I\notin \calS$
has a set of  independent generators $\calS=\langle \boldsymbol{\tau}_1,\ldots,\boldsymbol{\tau}_k\rangle$
such that 
\begin{equation}
\label{U}
\boldsymbol{\tau}_i=U \sigma^x_i U^\dag, \qquad i=1,\ldots,k
\end{equation}
for some Clifford unitary operator $U\in \calC_M$. 
Furthermore, the generators $\boldsymbol{\tau}_1,\ldots,\boldsymbol{\tau}_k$ 
can be constructed efficiently using the standard stabilizer formalism~\cite{gottesman1997stabilizer}, see   below.
Define a transformed Hamiltonian 
\begin{equation}
\label{Hsim'}
H_{sim}'=U^\dag H_{sim} U = \sum_{j=1}^r h_j \boldsymbol{\eta}_j, \qquad \boldsymbol{\eta}_j\equiv U^\dag
\boldsymbol{\sigma}_j U\in \cP_M.
\end{equation}
Clearly, $H_{sim}'$ and $H_{sim}$ have the same eigenvalues.
Furthermore, since $[\boldsymbol{\tau}_i,\boldsymbol{\sigma}_j]=0$
for all $i,j$, one infers that $[\sigma^x_i,\boldsymbol{\eta}_j]=0$ for all $i,j$.
In other words,  all terms in
$H_{sim}'$ commute with all $X$-type Pauli operators on the first $k$ qubits.
This is possible only if each term $\boldsymbol{\eta}_j$ that appears in $H_{sim}'$
acts on the first $k$ qubits by $I$ or $\sigma^x$. 
When looking for the ground energy of $H_{sim}'$ in the variational approach
one can replace Pauli operators $\sigma^x_1,\ldots,\sigma^x_k$ by their eigenvalues
$\pm 1$ and remove the first $k$ qubits from the simulator system. 
Although the Clifford transformation $U$ does not change the number of Pauli terms
in the simulator Hamiltonian, it can in principle increase their weight.
Hence, in some cases, the Pauli terms $\boldsymbol{\eta}_j$ in $H_{sim}'$ may be 
supported on a larger number of qubits compared with the Pauli terms $\boldsymbol{\sigma}_j$
in $H_{sim}$.

We now describe an efficient procedure to find generators of the symmetry group $\calS$.
We shall parameterize Pauli operators $\boldsymbol{\sigma}\in \calP_M$ by 
binary strings $(a_x|a_z)$ where $a_x,a_z\in \{0,1\}^M$ such that 
\[
\boldsymbol{\sigma}(a_x|a_z) =e^{i\phi}  \prod_{i\in a_x} \sigma^x_i \cdot \prod_{j\in a_z} \sigma^z_j
\]
for some phase factor $e^{i\phi}$ that we shall ignore. 
Then
\[
\boldsymbol{\sigma}(a_x|a_z)\boldsymbol{\sigma}(b_x|b_z)=(-1)^{a_x\cdot b_z + a_z\cdot b_x} 
\boldsymbol{\sigma}(b_x|b_z)\boldsymbol{\sigma}(a_x|a_z).
\] 
The set of Pauli operators $\boldsymbol{\sigma}_1,\ldots,\boldsymbol{\sigma}_r$
that appear in $H_{sim}$ can be represented by a binary matrix 
\[
G=\left[ \begin{array}{c}
G_x \\
\hline
G_z \\
\end{array} \right]
\]
of size $(2M)\times r$
such that $\boldsymbol{\sigma}_j=\boldsymbol{\sigma}(G_x^j|G_z^j)$, where $G_x^j$ and $G_z^j$
is the $j$-th column of $G_x$ and $G_z$ respectively.
The matrix $G$ is analogous to the generator matrix of stabilizer codes,
and we will stick to the same notational conventions. 
A Pauli operator $\boldsymbol{\sigma}(b_x|b_z)$ commutes with each term
in $H_{sim}$ iff $G^T\cdot (b_z|b_x)=0$ where $(b_z|b_x)$ is considered as a column vector
and $G^T$ is $G$ transposed. 
It will be convenient to define a parity check matrix
\begin{equation}
\label{GtoE}
E=\left[ \begin{array}{c|c}
E_x & E_z \\
\end{array} \right],
\qquad E_x\equiv (G_z)^T, \quad E_z\equiv (G_x)^T.
\end{equation}
The matrix $E$ has size $r\times (2M)$.
We conclude that any element of the symmetry group $\calS$ must have a form
$\boldsymbol{\sigma}(b_x|b_z)$ for some $(b_x|b_z)\in \mbox{ker}(E)$.

Let $b^1,\ldots,b^d\in \{0,1\}^{2M}$ be a set of linearly independent vectors that form 
a basis of $\mbox{ker}(E)$. We shall write $b^i=(b_x^i|b_z^i)$. 
Let us choose the symmetry group $\calS$  as any  maximal abelian subgroup
of the group generated by the Pauli operators $\boldsymbol{\sigma}(b^1),\ldots,\boldsymbol{\sigma}(b^d)$.
Such subgroup $\calS$ can be constructed, for example, by applying 
a symplectic version of the Gram-Schmidt orthogonalization 
to the set of basis vectors $b^1,\ldots,b^d\in \{0,1\}^{2M}$, see~\cite{gottesman1997stabilizer}.
It gives a maximal set of linearly independent vectors
$g^1,\ldots,g^k\in \mathrm{span}(b^1,\ldots,b^d)$ such that 
\[
g^i_x \cdot g^j_z + g^i_z \cdot g^j_x =0{\pmod 2}, \quad \forall \quad 1\le i,j\le k.
\]
Then generators of the symmetry group $\calS$ can be chosen as 
$\boldsymbol{\tau}_i=\boldsymbol{\sigma}(g^i_x|g^i_z)$
and $\calS=\langle \boldsymbol{\tau}_1,\ldots,\boldsymbol{\tau}_k\rangle$.

Let us comment on how to choose the Clifford transformation $U$ satisfying Eq.~(\ref{U}).
In all examples considered below the symmetry generators $\boldsymbol{\tau}_1,\ldots,\boldsymbol{\tau}_k$
are $Z$-type Pauli operators. 
In addition, we can choose a subset of qubits $q(1),\ldots,q(k)$ such that 
$\sigma^x_{q(i)}$ anti-commutes with $\boldsymbol{\tau}_i$
and commutes with $\boldsymbol{\tau}_j$ for all $j\ne i$, that is,
\begin{equation}
\label{xpauli}
\sigma^x_{q(i)} \boldsymbol{\tau}_j=(-1)^{\delta_{i,j}}\, \boldsymbol{\tau}_j\sigma^x_{q(i)}.
\end{equation}
Define unitary Clifford operators
\begin{equation}
\label{clifford}
U_i=\frac1{\sqrt{2}} ( \sigma^x_{q(i)} + \boldsymbol{\tau}_i),  \qquad i=1,\ldots,k.
\end{equation}
Using the commutation rules Eq.~(\ref{xpauli})
one can check that  
\[
U_i^2=I, \quad  U_i \sigma^x_{q(i)} U_i^\dag=\boldsymbol{\tau}_i,
\quad  \mbox{and} \quad 
U_j \sigma^x_{q(i)} U_j^\dag= \sigma^x_i
\]
for all $i\ne j$. 
Furthermore,  $U_1,\ldots,U_k$ pairwise commute. 
Thus a Clifford transformation $U$ satisfying Eq.~(\ref{U}) can be chosen as
$U=U_1 U_2\cdots U_kW$, where $W$ is a permutation of qubits
that maps qubits $1,\ldots,k$ to qubits $q(1),\ldots,q(k)$. Below we shall often ignore
the permutation $W$. 

To estimate utility of the above qubit reduction scheme in 
realistic examples, we have first applied the systematic search for symmetries to the $\textrm{H}_2$ Hamiltonian, reporting all the steps of the procedure in Appendix~\ref{app:H2}. We have then performed the symmetry search on a variety of small molecules, reported in Table~\ref{tablemol}. The one and two-body molecular integrals, i.e. the coefficients in Eq.~(\ref{Htgt}), for the molecules reported are obtained using the PyQuante open-source suite (version 1.6.0)~\cite{PyQuante}, in the chemists notation, using a STO-3G basis. All internuclear distances between different atoms are set to $1~\AA$, except for $\textrm{NH}_3$, and the angle in the tri-atomic molecules is set to $100\degree$ for $\textrm{H}_2\textrm{O}$ and $180\degree$ for $\textrm{BeH}_2$.  The geometry for $\textrm{NH}_3$ is the equilibrium geometry reported in the CCCBDB NIST archive~\cite{NIST} for a STO-3G basis using the configuration interaction method with single and double excitations. The second-quantized fermionic 
Hamiltonians are first symbolically mapped to qubit Hamiltonians using  the
Jordan-Wigner, the parity, and the binary tree encodings as detailed in 
Appendix~\ref{app:StandardMaps}, after which equal Pauli strings are recognized and 
simplified~\footnote{Ancillary data files that contain coefficients of the second-quantized  fermionic Hamiltonians
and the corresponding qubit Hamiltonians accompany the arXiv version of this paper.}.
The parity check matrix  $E$ defined in Eq.~(\ref{GtoE}) is built for every Hamiltonian
and the kernel of $E$ over the binary field  is computed.
The symmetry generators $\boldsymbol{\tau}_i$ together with single-qubit Pauli operators $\sigma^x_{q(i)}$
that obey the commutation rules Eq.~(\ref{xpauli})
and reported in Table~\ref{tablemol}.
As expected from Section~\ref{sec:parity}, using the parity encoding 
one always finds single-qubit symmetries $\sigma^z_j$ at 
qubits $j=M/2$ and $j=M$.
The same single-qubit symmetries at qubits $M/2,M$ are found for the binary tree encoding
 when the number of orbitals is a power of $2$, as in the case of $\textrm{NH}_3$, where $M=2^4$.

\begin{table*}[t]

\begin{tabular}{|c|c|c|c|c|}
\hline
& Spin Orbitals $M$ &Jordan-Wigner & Parity & Binary Tree\\
\hline
\hline

 \multirow{4}{*}{$\textrm{LiH}$} &\multirow{4}{*}{$12 $}		& 		$(\sigma^x_1,\sigma^z_1\sigma^z_2\sigma^z_3\sigma^z_6\sigma^z_{10}\sigma^z_{11} )$	 	&		$(\sigma^x_6,\sigma^z_6 )$ 		&		 $ (\sigma^x_4,\sigma^z_4 \sigma^z_6)$\\
 
&& $ (\sigma^x_4,\sigma^z_4 \sigma^z_{10} ) $		 &		 $ (\sigma^x_3,\sigma^z_3 \sigma^z_5 \sigma^z_9 \sigma^z_{11})$		  &		 $(\sigma^x_2,\sigma^z_2\sigma^z_3\sigma^z_6\sigma^z_9\sigma^z_{10})$ \\
 
&& $ (\sigma^x_5, \sigma^z_5 \sigma^z_{11}) $		&		 $(\sigma^x_4,\sigma^z_4 \sigma^z_5 \sigma^z_{10} \sigma^z_{11} )$		 &		$(\sigma^x_5,\sigma^z_5 \sigma^z_{11} )$\\

&& $ (\sigma^x_7,\sigma^z_7 \sigma^z_8 \sigma^z_9 \sigma^z_10 \sigma^z_{11} )$		 &		 $(\sigma^x_{12},\sigma^z_{12})$		&		$(\sigma^x_8,\sigma^z_8\sigma^z_{12})$\\
  
 \hline
  \multirow{4}{*}{$\textrm{BeH}_{2}$} 		&		 \multirow{4}{*}{14} 		&		 $(\sigma^x_1,\sigma^z_1\sigma^z_2\sigma^z_3\sigma^z_6\sigma^z_7\sigma^z_{11}\sigma^z_{12})$		 & 		$(\sigma^x_7,\sigma^z_7)$ 		&		 $(\sigma^x_4,\sigma^z_4\sigma^z_6\sigma^z_7)$\\
  
  && $(\sigma^x_4,\sigma^z_4\sigma^z_{11})$ 		&		 $(\sigma^x_3,\sigma^z_3 \sigma^z_5 \sigma^z_{10} \sigma^z_{12})$ 		&		 $(\sigma^x_2,\sigma^z_2\sigma^z_3\sigma^z_6\sigma^z_7\sigma^z_{11})$\\
  
  && $(\sigma^x_5,\sigma^z_5\sigma^z_{12})$ 		&		 $(\sigma^x_4,\sigma^z_4\sigma^z_5\sigma^z_{11}\sigma^z_{12})$		 &		 $(\sigma^x_5,\sigma^z_5 \sigma^z_{10} \sigma^z_{11}\sigma^z_{12})$\\
  
  && $(\sigma^x_8,\sigma^z_8\sigma^z_9\sigma^z_{10}\sigma^z_{11}\sigma^z_{12}\sigma^z_{13}\sigma^z_{14})$ 		&		 $(\sigma^x_{14},\sigma^z_{14})$		 &		 $(\sigma^x_8,\sigma^z_8\sigma^z_{12}\sigma^z_{14})$ \\
  
  \hline
  \multirow{3}{*}{$\textrm{H}_{2}\textrm{O}$} 		&		 \multirow{3}{*}{14} 		&		 $(\sigma^x_1,\sigma^z_1\sigma^z_2\sigma^z_3\sigma^z_5\sigma^z_6\sigma^z_7\sigma^z_{11})$		 &		$(\sigma^x_7,\sigma^z_7)$ 		&		 $(\sigma^x_4,\sigma^z_4\sigma^z_6\sigma^z_7)$\\
  
  &&  $(\sigma^x_4,\sigma^z_4 \sigma^z_{11})$ 		&		 $(\sigma^x_3,\sigma^z_3\sigma^z_4\sigma^z_{10}\sigma^z_{11})$ 		&		 $(\sigma^x_2,\sigma^z_2\sigma^z_3\sigma^z_6\sigma^z_7\sigma^z_{11})$ \\
  
  && $(\sigma^x_8,\sigma^z_8\sigma^z_9\sigma^z_{10}\sigma^z_{11}\sigma^z_{12}\sigma^z_{13}\sigma^z_{14})$		 &		 $(\sigma^x_{14},\sigma^z_{14})$		 &		 $(\sigma^x_8,\sigma^z_8\sigma^z_{12}\sigma^z_{14})$ \\
  
  \hline
 \multirow{2}{*}{$\textrm{NH}_{3}$} 		&		 \multirow{2}{*}{16}
 & 		$(\sigma^x_1,\sigma^z_1\sigma^z_2\sigma^z_3\sigma^z_4\sigma^z_5\sigma^z_6\sigma^z_7\sigma^z_8)$ 		&		 $(\sigma^x_8,\sigma^z_8)$ 		&		 $(\sigma^x_8,\sigma^z_8)$ \\ 
 
 && $(\sigma^x_9,\sigma^z_9\sigma^z_{10}\sigma^z_{11}\sigma^z_{12}\sigma^z_{13}\sigma^z_{14}\sigma^z_{15}\sigma^z_{16})$ 		&		 $(\sigma^x_{16},\sigma^z_{16})$		 &		 $(\sigma^x_{16},\sigma^z_{16})$ \\
 
 \hline 
 \multirow{4}{*}{$\textrm{HCl}$} 		&		 \multirow{4}{*}{20}		&		 $(\sigma^x_1,\sigma^z_1\sigma^z_2\sigma^z_3\sigma^z_6\sigma^z_7\sigma^z_{10}\sigma^z_{14}\sigma^z_{15}\sigma^z_{18}\sigma^z_{19})$ 		& 
 $(\sigma^x_{10},\sigma^z_{10})$  & $(\sigma^x_8,\sigma^z_{8}\sigma^z_{10})$\\
 
 && $(\sigma^x_4,\sigma^z_4\sigma^z_8\sigma^z_{14}\sigma^z_{18})$ 		& 		$(\sigma^x_3,\sigma^z_3\sigma^z_5\sigma^z_7\sigma^z_9\sigma^z_{13}\sigma^z_{15}\sigma^z_{17}\sigma^z_{19})$ 		&		 $(\sigma^x_2,\sigma^z_2\sigma^z_3\sigma^z_6\sigma^z_7\sigma^z_{10}\sigma^z_{13}\sigma^z_{14}\sigma^z_{17}\sigma^z_{18})$ \\
 
 && $(\sigma^x_5,\sigma^z_5\sigma^z_9\sigma^z_{15}\sigma^z_{19})$ 		&		$(\sigma^x_4,\sigma^z_4\sigma^z_5\sigma^z_8\sigma^z_9\sigma^z_{14}\sigma^z_{15}\sigma^z_{18}\sigma^z_{19})$		 &		$(\sigma^x_5,\sigma^z_5\sigma^z_9\sigma^z_{15}\sigma^z_{19})$ \\
 
 && $(\sigma^x_{11},\sigma^z_{11}\sigma^z_{12}\sigma^z_{13}\sigma^z_{14}\sigma^z_{15}\sigma^z_{16}\sigma^z_{17}\sigma^z_{18}\sigma^z_{19}\sigma^z_{20})$ 		&
$(\sigma^x_{20}, \sigma^z_{20})$ 		&		 $(\sigma^x_{16},\sigma^z_{16}\sigma^z_{20})$\\

 \hline
 
\end{tabular}

\caption{\label{tablemol} List of symmetry generators $\boldsymbol{\tau}_i$
found for a set of small molecules, reported together with the corresponding single-qubit operators
$\sigma^x_{q(i)}$ that obey commutation rules Eq.~(\ref{xpauli}).
The table follows  the notation $(\sigma^x_{q(i)}, \boldsymbol{\tau}_i)$. The molecular Hamiltonians for the molecules reported are computed in the STO-3G basis, using one and two-body integrals obtained from the PyQuante software (version 1.6.0). The internuclear distances between different atoms are set to $1~\AA$, and the angle in the tri-atomic molecules was set to $100\degree$ for $\textrm{H}_2\textrm{O}$ and $180\degree$ for $\textrm{BeH}_2$. The geometry for $\textrm{NH}_3$ is the equilibrium one for the basis set considered.}
\end{table*}

\appendix

\section{Standard fermionic mappings}
\label{app:StandardMaps}
In this section we recall the definitions for known mappings from a fermionic Fock space to qubits. The  Jordan-Wigner transformation~\cite{Wigner1928} maps $M$ fermions on to $M$ ordered qubits by assigning to the value of the $j$-th qubit the occupation of the $j$-th fermionic mode, and stores the parity information on the occupation of the modes preceiding the index $j$ with a $Z$ check on the corresponding qubits. It is defined as a correspondence between fermionic creation and annihilation operators and qubit operators,
\begin{equation}
\label{JWMap}
a_j \rightarrow  \left(\prod_{i=1}^{j-1}\sigma^z_i \right)\sigma^+_j  \Sp \mbox{and} \Sp  a^\dag_i \rightarrow  \left(\prod_{i=1}^{j-1}\sigma^z_i\right)\sigma^-_j.
\end{equation}
We have used the usual definitions $\sigma^+=(\sigma^x+i\sigma^y)/2$ and $\sigma^-=(\sigma^x-i\sigma^y)/2$. The action of the JW operators in Eq.~(\ref{JWMap}) on a qubit state vector can be seen as a flip on the $i$th bit, carrying a sign obtained by counting the number of $1$-bits in the subset with index less than $i$.  This transformation leads to Pauli operators that are supported on $\cO(M)$ qubits.

The parity mapping~\cite{BK2002,seeley2012bravyi} encodes in the $j$-th qubit the information of the parity of the $j$-th fermionic mode and the ones preceiding it, being dual to the Jordan-Wigner transformation. It reads
\begin{equation}
\begin{split}
a_j\rightarrow&\frac{1}{2}\left[ \sigma^z_{j-1} \sigma^x_j \prod_{i=j+1}^M \sigma^x_i+i\sigma^y_j \prod_{i=j+1}^M \sigma^x_i\right]\\
a^\dag_j\rightarrow&\frac{1}{2}\left[ \sigma^z_{j-1} \sigma^x_j \prod_{i=j+1}^M \sigma^x_i-i\sigma^y_j \prod_{i=j+1}^M \sigma^x_i\right].\\
\end{split}
\end{equation}
The parity mapping, as the Jordan-Wigner map, contains linear strings of operators, and therefore can map local fermionic operators into Pauli strings supported on $\cO(M)$ qubits.
The binary tree mapping~\cite{BK2002} improves on this linear scaling, with $\cO(\log(M))$-local terms.  It has  a compact definition in the form~\cite{seeley2012bravyi}
\begin{equation}
\label{BKmapped}
\begin{split}
a_j&\rightarrow \frac{1}{2} \prod_{i\in U(j)}  \sigma^x_i\times\left[\sigma^x_j\prod_{i\in P(j)} \sigma^z_i+i\sigma^y_j\prod_{i\in R(j)} \sigma^z_i\right]\\
a^\dag_j&\rightarrow \frac{1}{2} \prod_{i\in U(j)}  \sigma^x_i\times\left[\sigma^x_j\prod_{i\in P(j)} \sigma^z_i-i\sigma^y_j\prod_{i\in R(j)} \sigma^z_i\right],
\end{split}
\end{equation}
The update ($U(j)$), parity ($P(j)$) and flip ($F(j)$) sets can be obtained systematically from the partial order on binary strings~\cite{BK2002} or, equivalently, from the recursive matrices that maps fermionic occupation into qubits~\cite{seeley2012bravyi,Tranter15}. The remainder set $R(j)=P(j) \setminus F(j)$ is obtained from the set difference of the parity and the flip sets. We report recursive formulas in Ref.~\cite{Tranter15} to find the update, parity and flip set in a system of $j=1,2,...,M$ qubits,
\begin{eqnarray}
\nonumber
&U_M(j)&=\begin{cases}
\left\{U_{M/2}(j-1)+1, M\right\},\; j<\frac{M+2}{2}\\
\left\{U_{M/2}(j-1-\frac{M}{2})+\frac{M+2}{2}\right\},\; j\geq \frac{M+2}{2}
\end{cases}\\
\nonumber
&P_M(j)&=\begin{cases}
P_{M/2}(j-1)+1,\; j<\frac{M+2}{2}\\
\left\{P_{M/2}(j-1-\frac{M}{2})+\frac{M+2}{2},\frac{M}{2}\right\},\; j\geq \frac{M+2}{2}
\end{cases}\\
&F_M(j)&=\begin{cases}
F_{M/2}(j-1),\; j<\frac{M+2}{2}\\
\nonumber
\{F_{M/2}(j-1-\frac{M}{2})+\frac{M+2}{2}\},\; \frac{M+2}{2}\leq j< M\\
\{F_{M/2}\left(j-1-\frac{M}{2}\right)+\frac{M+2}{2}, \frac{M}{2}\},\;j= M.
\end{cases}
\end{eqnarray}

\section{The hydrogen molecule}
\label{app:H2}

We consider a simple example where tapering of qubits can be followed step by step. We consider the $\textrm{H}_2$ Hamiltonian derived in Ref.~\cite{whitfield2011simulation} by working with four fermionic modes in a minimal basis. Note that the $\textrm{H}_2$ Hamiltonian considered here does not have the orbital ordering in which the first $2$ orbitals are the spin-up ones. 
In Ref.~\cite{o2015scalable} it was already observed that two qubits could be removed after the binary tree encoding.
Here we apply the method of Section~\ref{sec:multiple}
with the standard Jordan-Wigner  encoding
 to show that one can in fact remove three qubits from the four qubit Hamiltonian. 

The qubit simulator Hamiltonian $H_{sim}$ for the hydrogen molecule 
has the form Eq.~(\ref{Hpauli})
where the number of qubits is $M=4$, the number of Pauli terms is $r=14$ and  all
Pauli operators $\boldsymbol{\sigma}_j\in \calP_4$ that appear in $H_{sim}$ are listed 
in the following table~\cite{whitfield2011simulation}:
\bq
\label{H2Pauli}
\begin{array}{l | l | l | l}
\sigma^z_1 & \sigma^z_2 & \sigma^z_3 & \sigma^z_4 \\ \hline
\sigma^z_1  \sigma^z_2  & \sigma^z_1  \sigma^z_3  & \sigma^z_1  \sigma^z_4 & \\ \hline
\sigma^z_2  \sigma^z_3  & \sigma^z_2  \sigma^z_4 & \sigma^z_3  \sigma^z_4 &    \\ \hline
\sigma^y_1  \sigma^y_2 \sigma^x_3  \sigma^x_4 &  \sigma^x_1  \sigma^y_2 \sigma^y_3  \sigma^x_4 
& \sigma^y_1  \sigma^x_2 \sigma^x_3  \sigma^y_4 & \sigma^x_1  \sigma^x_2 \sigma^y_3  \sigma^y_4  
\end{array}
\eq
These operators can be mapped to the matrix $G$, by constructing their binary representation and arranging each Pauli as a column vector in $G$, which is now a $ 8 \times 14$ binary matrix. For instance the Pauli operator $\sigma^y_1  \sigma^y_2 \sigma^x_3  \sigma^x_4$ is represented by $( 1 1 1 1 | 1 1 0 0 )$, whereas  for $\sigma^z_1  \sigma^z_3$ we obtain $(0 0 0 0 | 1 0 1 0)$. From the matrix $G$ we can now construct
the parity check matrix  $E$ defined  in Eq.~(\ref{GtoE}). 
Let $\tilde{E}$ be the row-echelon form of $E$ obtained by performing 
the  Gaussian elimination. After the removal of trivial rows we get
\bq
\tilde{E} = \left(\begin{array}{cccc | cccc}
1 & 0  & 0 & 0 & 0 & 0  & 0 & 0 \\
0 & 1  & 0 & 0 & 0 & 0  & 0 & 0 \\
0 & 0  & 1 & 0 & 0 & 0  & 0 & 0 \\ 
0 & 0  & 0 & 1 & 0 & 0  & 0 & 0 \\ 
0 & 0  & 0 & 0 & 1 & 1  & 1 & 1
\end{array} \right).
\eq  
We choose three linearly independent vectors 
\[
(0 0 0 0 | 1 1 0 0), \quad  (0 0 0 0 | 1 0 1 0), \quad  (0 0 0 0 | 1 0 0 1)
\]
that span the kernel of $\tilde{E}$ (which coincides with the kernel of $E$).
These vectors give rise to three symmetry generators $\boldsymbol{\tau}_1 = \sigma^z_1 \sigma^z_2$, $\boldsymbol{\tau}_2 = \sigma^z_1 \sigma^z_3$ and $\boldsymbol{\tau}_3 = \sigma^z_1 \sigma^z_4$ respectively. Next let us choose  single qubit Pauli operators $\sigma^x_{q(i)}$ that obey commutation rules Eq.~(\ref{xpauli}). In this example we can choose
$\sigma^x_{q(i)}=\sigma^x_{i+1}$ for $i=1,2,3$. We use the pairs $(\sigma^x_{i+1},\boldsymbol{\tau}_i)$ to construct the mutually commuting Clifford operators Eq.~(\ref{clifford}), that is,
\bq
&&U_1 = \frac{1}{\sqrt{2}} \left( \sigma^x_2 + \sigma^z_1 \sigma_2^z \right), \;\; U_2 = \frac{1}{\sqrt{2}} \left( \sigma^x_3+ \sigma^z_1 \sigma_3^z \right) \no
&& \mbox{and} \Sp U_3 = \frac{1}{\sqrt{2}} \left( \sigma^x_4 + \sigma^z_1 \sigma_4^z \right).
\eq
We now conjugate every Pauli operator $\boldsymbol{\sigma}_j$ from  the table Eq.~(\ref{H2Pauli}) by the Clifford operator $U=U_1U_2U_3$.
The resulting $14$ Pauli operators $\boldsymbol{\eta}_j=U^\dag\boldsymbol{\sigma}_j U$ are given by the following table:
\bq
\label{transformedH2}
\begin{array}{l | l | l | l}
\sigma^z_1 & \sigma^z_1\mgr{\sigma^x_2} &\sigma^z_1 \mgr{ \sigma^x_3} & \sigma^z_1 \mgr{\sigma^x_4} \\ \hline
\mgr{\sigma^x_2}  &   \mgr{\sigma^x_3}  &  \mgr{ \sigma^x_4} & \\ \hline
\mgr{\sigma^x_2  \sigma^x_3}  & \mgr{\sigma^x_2  \sigma^x_4} &\mgr{ \sigma^x_3  \sigma^x_4} &    \\ \hline
\sigma^x_1 \mgr{\sigma^x_3  \sigma^x_4} &  \sigma^x_1 \mgr{\sigma^x_4} & \sigma^x_1  \mgr{\sigma^x_2 \sigma^x_3} & \sigma^x_1 \mgr{\sigma^x_2}  
\end{array}
\eq
This transformation 
gives a new simulator Hamiltonian $H_{sim}'=\sum_{j=1}^{14} h_j\boldsymbol{\eta}_j$ defined in Eq.~(\ref{Hsim'}). We observe that the Pauli operators in table (\ref{transformedH2}) act on qubits $2,3,4$ by either $\sigma^x$ or $I$. Hence, these qubits can be removed from $H_{sim}'$ and the Pauli matrices  $\sigma^x_2,\sigma^x_3,\sigma^x_4$ can be replaced by their eigenvalues $\pm 1$. Hence, the hydrogen Hamiltonian reduces to a trivial single qubit problem.

\section{Decoding algorithm}
\label{app:decoding}

Here we explain how to compute  the set $A^{-1}s$ defined in Eq.~(\ref{invA}).
Below we use notations and terminology of Section~\ref{sec:LDPC}.
Let $W(M,N)$ be the set of all $M$-bit strings with weight $N$
and $A$ be a fixed $N$-injective matrix of size $Q\times M$. 
One can easily check that $A$ is $N$-injective iff 
\begin{equation}
\label{inject1}
\ker{(A)}\cap W(M,2K) = \emptyset \quad \mbox{for all} \quad 1\le K\le N.
\end{equation}
Here $\ker{(A)}=\{x \in \{0,1\}^M\, : \, Ax=0\}$. As a corollary of Eq.~(\ref{inject1}),
one infers that $A$ is $K$-injective for all $K\le M$. 
Decompose $N=N_1+N_2$, where
$N_{1,2}=N/2$ for even $N$ and $N_{1,2}=(N\pm 1)/2$ for odd $N$.
For each $i=1,2$ 
let $T_i$ be a lookup table that stores
syndromes $t=Au$ for each $u\in W(M,N_i)$. 
The entries of $T_i$ are sorted in the lexicographic order. 
Let $U_i$ be a lookup table that maps
each entry $t\in T_i$ to a string $u\in W(M,N_i)$
such that $t=Au$. Note that $u$ as above is unique since  $A$
is $N_i$-injective. 
The tables $T_i,U_i$ can be computed offline since they depend only on $A$.

Suppose first that $s=Ax$ for some $x\in W(M,N)$, that is, $A^{-1}s=\{x\}$. Consider any decomposition
$x=u_1\oplus u_2$ with $u_i\in W(M,N_i)$ and let $t_i=Au_i$.
Then the tables $T_1$ and $T_2$ must contain entries $t_1$ and $t_1\oplus s$
respectively. For each $t_1\in T_1$ let us check whether
$T_2$ contains $t_2=t_1\oplus s$.  This can be done
in time $O(|T_1| \log{|T_2|})$ using the binary search since the table $T_2$ is sorted. 
Suppose we found $t_1,t_2$ as above.
Use the tables $U_i$ to find $u_i$ such that $t_i=Au_i$. 
Then $A(u_1\oplus u_2)=s$ and 
\[
|u_1\oplus u_2|=|u_1| + |u_2| -2|u_1\cap u_2|=N-2|u_1\cap u_2|.
\]
It follows that $A(x\oplus u_1\oplus u_2)=0$ and $x\oplus u_1\oplus u_2$ has even
weight between $2$ and $2N$. From Eq.~(\ref{inject1}) one infers that
$x=u_1\oplus u_2$ and we are done. In the remaining case, if 
a pair $t_1,t_2$ as above is not found, 
we infer that $Ax=s$ has no solutions with $x\in W(M,N)$,
that is, $A^{-1}s=\emptyset$.
The above algorithm is practical 
for medium size systems, say,  $M\le 50$. Indeed, since $M\le N/2$,
the tables $T_i,U_i$ have size at most ${M\choose M/4}\approx 4\times 10^{11}$ 
for $M=50$. This would require roughly 1TB of memory.

\textit{Acknowledgments}: We are grateful to 
Nikolaj Moll, 
Jed Pitera,
Gavin Jones,
and Julia Rice for helpful discussions. 
We acknowledge support from the IBM Research Frontiers Institute.


\end{document}